\documentstyle[preprint,epsfig,aps]{revtex}
\begin{document}
\draft
\tighten
%%%%%%%%%%%%%%%%%%%%%%%%%%
%%%%%%%%%%%%%%%%%% defs %%%%%%%%%%%%%%%%%%
\newcommand{\newc}{\newcommand}
\newc{\be}{\begin{equation}}
\newc{\ee}{\end{equation}}
\newc{\bea}{\begin{eqnarray}}
\newc{\eea}{\end{eqnarray}}
\newc{\Rslash}{\not R}
% superfields:
\newc{\superhu}{\hat H_u}
\newc{\superhd}{\hat H_d}
\newc{\superl}{\hat L}		
\newc{\superr}{\hat R}
\newc{\superec}{\hat {e}^c}
\newc{\superq}{\hat Q}
\newc{\superu}{\hat U}
\newc{\superd}{\hat D}
\newc{\superuc}{\hat {u}^c}
\newc{\superdc}{\hat {d}^c}
\newc{\Lsoft}{{\cal L}_{\rm soft}}
\newc{\hc}{{\rm H.c.}}
% scalar fields:
\newc{\tildeQ}{\widetilde Q}
\newc{\tildeU}{\widetilde U}
\newc{\tildeD}{\widetilde D}
\newc{\tildeL}{\widetilde L}
\newc{\tildeuc}{\widetilde {u}^c}
\newc{\tildedc}{\widetilde {d}^c}
\newc{\tildeec}{\widetilde {e}^c}
\newc{\tildenu}{\widetilde \nu}
\newc{\onehalf}{\frac{1}{2}}
\newc{\gluino}{{\tilde g}}
\newc{\mgluino}{m_{\gluino}}
\newc{\mzero}{m_0}
\newc{\mz}{m_Z}
\newc{\mw}{m_W}
\newc{\alphas}{\alpha_s}
\newc{\tanb}{\tan\beta}
\newc{\hinot}{\widetilde H^0_2}
\newc{\hinob}{\widetilde H^0_1}
\newc{\bino}{\widetilde B^0}
\newc{\wino}{{\widetilde W^0_3}}
\newc{\msl}{m_{\widetilde l}}
\newc{\msel}{m_{\tilde e}}
\newc{\msq}{m_{\tilde q}}
\newc{\msf}{m_{\tilde f}}
\newc{\mev}{{\rm\,MeV}}
\newc{\gev}{{\rm\,GeV}}
\newc{\tev}{{\rm\,TeV}}
\def\mat#1#2#3#4{\left(\begin{array}{cc}#1&#2\\#3&#4\end{array}\right)}
\def\vect#1#2{\left(\begin{array}{c}#1\\#2\end{array}\right)}
\def\bi{\beta}
\newc{\swsq}{\sin^2\theta_W}
\def\mHu{m_{H_u}}
\def\mHd{m_{H_d}}
\def\rp{$R_p \hspace{-1em}/\;\:$}

\def\NPB#1#2#3{Nucl. Phys.  {\bf B#1}, #3 (19#2)}
\def\PLB#1#2#3{Phys. Lett. B {\bf#1}, #3 (19#2)}
\def\PLBold#1#2#3{Phys. Lett. {\bf#1B}, #3 (19#2)}
\def\PRD#1#2#3{Phys. Rev. D {\bf #1}, #3 (19#2)}
\def\PRL#1#2#3{Phys. Rev. Lett. {\bf#1}, #3 (19#2)}
\def\PRT#1#2#3{Phys. Rep. {\bf#1}, #3 (19#2)}
\def\PPNP#1#2#3{Prog. Part. Nucl. Phys {\bf#1}, #3 (19#2)}
\def\ARAA#1#2#3{Ann. Rev. Astron. Astrophys. {\bf#1}, #3 (19#2)}
\def\ARNP#1#2#3{Ann. Rev. Nucl. Part. Sci. {\bf#1} #3 (19#2)}
\def\MODA#1#2#3{Mod. Phys. Lett. {\bf A#1}, #3 (19#2)}
\def\ZPC#1#2#3{Zeit. f\"ur Physik {\bf C#1}, #3 (19#2)}
\def\APJ#1#2#3{Ap. J. {\bf#1}, #3 (19#2)}

%%%%%%%%%%%%%%%%%% defs %%%%%%%%%%%%%%%%%%

\title{Grand unified theory constrained supersymmetry
       and neutrinoless double beta decay}

\author{Andrzej Wodecki, Wies{\l}aw A. Kami{\'n}ski} 

\address{ Department of Theoretical Physics,
Maria Curie--Sklodowska University,\\
PL-20 031 Lublin, Poland }

\author{Fedor \v Simkovic}

\address{Department of Nuclear Physics, Comenius University,\\
 Mlynsk\'a dolina F1, SK-842 15 Bratislava, Slovakia}

\date{\today}
\maketitle

%%%%%%%%%%%%%%%%%%%%%%%%%%%%%%%%%%%%%%%%%%%%%%%%%%%%%%%%%%%%%%%%%%
%%%%%%%%%%%              Abstract                         %%%%%%%%
%%%%%%%%%%%%%%%%%%%%%%%%%%%%%%%%%%%%%%%%%%%%%%%%%%%%%%%%%%%%%%%%%%
\begin{abstract}
We analyze the contributions to the neutrinoless double $\beta$ decay 
($0\nu\beta\beta$-decay) coming from the Grand Unified Theory (GUT) 
constrained Minimal Supersymmetric Standard Model (MSSM) 
with trilinear R--parity breaking. 
We discuss the importance of two-nucleon and pion-exchange realizations 
of the quark-level $0\nu\beta\beta$-decay transitions. In this context,
the questions of reliability of the calculated relevant nuclear 
matrix elements within the
Renormalized Quasiparticle Random Phase Approximation 
(pn-RQRPA) for several medium and heavy open-shell nuclei are addressed.
The importance of  gluino and neutralino 
contributions to $0\nu\beta\beta$-decay  is also analyzed. 
We review the present experiments and deduce limits on the trilinear 
R-parity breaking parameter $\lambda_{111}'$ from the non-observability
 of $0\nu\beta\beta$-decay for different
GUT constrained SUSY scenarios.
In addition, a detailed study of limits on the
MSSM parameter space coming from the $B \to  X_s \gamma$ processes 
by using the recent CLEO and OPAL results is performed.
Some studies in respect to the 
future $0\nu\beta\beta$-decay project GENIUS are also presented.

\end{abstract}

\pacs{12.60.Jv,11.30.Er,23.40.Bw}.

%%%%%%%%%%%%%%%%%%%%%%%%%%%%%%%%%%%%%%%%%%%%%%%%%%%%%%%%%%%%%%%%%%
%%%%%%%%%%%              Introduction                     %%%%%%%%
%%%%%%%%%%%%%%%%%%%%%%%%%%%%%%%%%%%%%%%%%%%%%%%%%%%%%%%%%%%%%%%%%%
\section{Introduction}

The neutrinoless double beta decay ($0\nu\beta\beta$-decay)
is forbidden in the Standard Model (SM) since it violates 
lepton number by two units ($\Delta L =2$). Therefore this decay 
is a sensitive 
probe for different aspects of physics beyond  SM.
(For recent reviews see e.g. \cite{FS98,SC98}). Many generalization of
the SM admit the violation of laws of the SM to a small extent.
In this context, non--observability of the $0\nu\beta\beta$-decay is 
used to constrain different extensions of the SM like
those with the left--right symmetry  \cite{doi83,moh98},
leptoquarks \cite{HKS96a}, R--parity violating
supersymmetric (\rp SUSY) models 
\cite{Moh86,Ver87,HKS96b,WKS97,FKSS97,FKS98a,FKS98b} and 
composite neutrinos \cite{Cab84,Pan97}.

The most widely discussed case in literature has been 
the upper bound on the light effective  electron neutrino 
Majorana mass $\langle m_\nu \rangle$ deduced from the 
experimental lower limit on the  half-life of $0\nu\beta\beta$-decay 
\cite{FS98,SC98}. Currently, the most restrictive limit on 
$\langle m_\nu \rangle$
is found from   $0\nu\beta\beta$-decay in $^{76}$Ge by the  
Heidelberg--Moscow collaboration \cite{bau97}: 
$\langle m_\nu \rangle \le 0.4-1.3$ eV \cite{FS98}.
The uncertainty of this parameter is due to the ambiguity of 
$0\nu\beta\beta$-decay nuclear matrix elements.
It is expected that the future double beta decay experiment 
GENIUS \cite{Hel97} based on  
1 tonn of enriched $^{76}$Ge would reach  the sensitivity for
$\langle m_\nu \rangle$ (0.01 - 0.001 eV).  

Besides the simplest and the best known  mechanism of lepton number 
violation based on the mixing of massive Majorana neutrinos advocated  
by different variants of the Grand Unified Theories (GUT) the R-parity 
violation proposed in the context of minimal supersymmetry
extensions of the SM  (MSSM) is becoming the most popular scenario
for lepton number violation (see e.g. reviews \cite{BAR98,BFK98}).
The R-parity is a multiplicative quantum number
defined as $R = (-1)^{3B+L+2S}$ with B, L and S 
being the baryon, the lepton and the spin quantum numbers of 
the particle, respectively. Thus, all the SM particles have R=+1, 
while their superpartners have R=-1. We note that the R-parity conservation,
which guarantee the baryon and lepton number conservation, 
is not required by gauge invariance or supersymmetry and might be 
broken explicitly or spontaneously at the Planck scale  \cite{rom92}. 

The R--parity can be broken by involving
the bilinear and trilinear terms in superpotential of the MSSM.
The bilinear terms generate non--zero vacuum expectation
value for the sneutrino field, which leads to the lepton number 
violating interactions based on the neutralino--neutrino 
and chargino--electron mixing \cite{FKS98b}. 
The trilinear terms represent the interactions, which 
violate directly the lepton number and lepton
flavor \cite{HKS96b,WKS97,FKSS97,FKS98a,FKS98b}.
Both ways influence the low--energy phenomenology and therefore
using some exotic processes like the $0\nu\beta\beta$-decay
one can impose limits on the parameters connected with the new physics.

Supersymmetric models with R-parity non-conservation (\rp MSSM)  
have been extensively discussed in the literature (see e.g. 
\cite{rbreaking,valle}) and were also used for the study of 
R-parity violating trilinear term contribution to the 
$0\nu\beta\beta$-decay by Mohapatra \cite{Moh86}  and Vergados \cite{Ver87}.
The first calculations were concentrated only on the conventional 
two-nucleon mode of $0\nu\beta\beta$-decay, which assumes
direct interactions between quarks of two-decaying neutrons
\cite{Moh86,Ver87,HKS96b,WKS97}. A detail study of this 
mechanism was carried out in Ref. \cite{HKS96b}. By using viable 
phenomenological assumptions about some of the fundamental 
parameters of the \rp MSSM (e.g. the ansatz of universal 
sparticle masses and that the lightest neutralino is
bino-like) it was found that the limit on the R-parity violating first 
generation Yukawa coupling $\lambda_{111}'$ derived from the 
observed absence of the $0\nu\beta\beta$-decay is more stringent than 
the corresponding limit expected from the forthcoming accelerator 
experiment at HERA. The authors end up with the conclusion that the gluino exchange 
R-parity violating mechanisms of the $0\nu\beta\beta$-decay 
dominates over the neutralino ones \cite{HKS96b}.

Another scenario for reduction of a number of supersymmetric parameters
associated with the limit on $\lambda_{111}'$ has been outlined
in Ref. \cite{WKS97}. The authors implemented 
relations among the weak scale values of all parameters
entering the superpotential and the soft SUSY breaking Lagrangian and 
their values at the GUT scale. The obtained results showed 
the importance of the neutralino exchange mechanism. Similar studies have
been performed also within gauge mediated SUSY breaking model
\cite{gauge1,gauge2}. We note that the  GUT's constrained SUSY  
scenarios have advantages of fewer degrees of freedom and usually 
predict more stringent limits on $\lambda_{111}'$ \cite{gauge1,gauge2}.

Recently, the dominance of the pion-exchange R-parity violating 
$0\nu\beta\beta$-decay 
mechanism (based on the double-pion exchange between the decaying neutrons)
over the conventional two-nucleon one was proven for $^{76}$Ge 
isotope \cite{FKSS97}. We note that the attention to the pion-exchange 
$0\nu\beta\beta$-decay mechanism was paid out  first by Pontecorvo
\cite{bruno} and that this mechanism  has been found less important in the
presence of light Majorana neutrinos and  more important, if heavy 
Majorana neutrinos are considered \cite{ver82}. 
There are two advantages, which favor pion-exchange R-parity violating
mode over the two-nucleon mode. First, the effective radius of the
two-nucleon R-parity violating interaction is small because
of exchange of heavy SUSY particles and therefore this mode is suppressed
by the nucleon-nucleon repulsion at short distances. On the other side
the pion-exchange mode leads to a long-range nuclear interaction,
which is significantly less sensitive to short--hand correlations effects,
as is mediated by light
particles. Secondly, the enhancement of the pion-exchange mode has an origin
in the bosonization of the $\pi^- -> \pi^+ + 2e^-$ vertex and is
associated with the pseudoscalar hadronic current structure  of the 
effective R-parity violating $0\nu\beta\beta$-decay Lagrangian on quark level
\cite{FKSS97,FKS98b}. The new pion-exchange mechanism 
of $0\nu\beta\beta$-decay offers more stringent limits on 
$\lambda_{111}'$ \cite{FKSS97}. 
The one- and two-pion exchange realization of this mode have been
discussed for experimentally interesting nuclear systems in 
Ref. \cite{FKS98b}. Both gluino and neutralino exchange mechanisms 
have been found of comparable importance within the phenomenological 
scenario of this mode.

The procedure leading to the final constraint on $\lambda_{111}'$
consists of two parts: In the first step, 
the relevant SUSY parameters are determined
within a proper SUSY scenario. Next, the reliable evaluation of
$0\nu\beta\beta$-decay nuclear matrix elements has to be performed.
Since the importance of the particle-particle residual interaction 
of the nuclear Hamiltonian for the description of open shell nuclear
system was discovered \cite{VZ86}, the proton-neutron Quasiparticle Random
Phase Approximation (pn-QRPA) has been widely chosen in the calculations
of the  nuclear double beta decay transitions 
\cite{FS98,SC98,CFT87,EVZ88,MK88,KS94,RFSK91,FKPV,StKam,PSVF96}. 
However, the extreme sensitivity of the calculated nuclear matrix 
elements as well as the collapse of the QRPA solution in the
physically acceptable region of particle-particle strength 
address some questions  about the predictive power of the obtained results
\cite{SSVP97,SPF98}. The renormalized pn-QRPA (pn-RQRPA) without 
\cite{TS95} and with proton-neutron pairing \cite{SSF96}, which take
into account the Pauli exclusion principle, do not collapse
and offer more reliable results in respect to the QRPA \cite{SSVP97,SPF98}. 
This method has been used also in the recent nuclear structure studies
of the $0\nu\beta\beta$-decay matrix elements 
\cite{FS98,FKSS97,FKS98a,FKS98b,SSVP97}. Some other extensions of RQRPA
can be also addressed  in this context 
\cite{WAK98}.

The goal of the present paper is to perform a comprehensive analysis
of the R-parity violating trilinear term contribution to 
$0\nu\beta\beta$-decay within the GUT constrained MSSM. 
We shall focus our attention on
a detailed pn-RQRPA study of the relative importance of the pion-exchange
and two-nucleon modes of this process for  experimentally 
interesting $0\nu\beta\beta$-decay nuclei. Our aim is also to discuss
the present  as well as possible future (imposed by GENIUS experiment)
limit on the trilinear breaking SUSY parameter $\lambda_{111}'$
and identify this limit with the gluino or the neutralino exchange mechanisms.
Moreover, all the calculations are performed within MSSM constrained by 
different processes, among which the most interesting are limits coming
from $B \to  X_s \gamma$ processes obtained
by using the recent CLEO and OPAL results.

The paper is arranged as follows. In the next section (section II) we 
introduce the basic elements of the MSSM with the explicit R-parity
breaking. Here we also discuss the main formulae relevant for the trilinear R-parity 
breaking contribution to $0\nu\beta\beta$-decay 
and the realistic pn--RQRPA nuclear structure method, which will be used for 
evaluation of nuclear matrix elements of interest. 
In section III we calculate the $0\nu\beta\beta$-decay 
matrix elements within the pn--RQRPA for experimentally interesting
isotopes and analyze their 
uncertainties in respect to details of the nuclear model.
In addition, we discuss the parameters of the MSSM
and the importance of
the gluino and the neutralino $0\nu\beta\beta$-decay mechanisms. 
The experimental constraints on  the $0\nu\beta\beta$-decay are then 
used to constrain
the first generation of lepton number violating Yukawa coupling 
constant  ${\acute{\lambda}}_{111}$ of supersymmetric particles.
We close with  a short summary and conclusions (section V).

%%%%%%%%%%%%%%%%%%%%%%%%%%%%%%%%%%%%%%%%%%%%%%%%%%%%%%%%%%%%%%%%%%
%%%%%%%%%%%                Theory                         %%%%%%%%
%%%%%%%%%%%%%%%%%%%%%%%%%%%%%%%%%%%%%%%%%%%%%%%%%%%%%%%%%%%%%%%%%%
\section{Theory}

%%%%%%%%%%%%%%%%%%%%%%%%%%%%%%%%%%%%%%%
\subsection{R-parity violating minimal supersymmetric standard model}
%%%%%%%%%%%%%%%%%%%%%%%%%%%%%%%%%%%%%%%

The MSSM  is based on the 
same gauge group as the SM, 
It is particle content required to 
implement supersymmetry in a
consistent way is the minimal one . 
It is described by the superpotential 
which in the case of R-parity non-conservation 
contains both the R-parity conserving and breaking parts
\be
W = W_0 + W_{\Rslash},
\label{eq:1}
\ee
where
\be
W_0 = h^U_{ij} \superq_i\superhu\superuc_j
           + h^D_{ij} \superq_i\superhd\superdc_j
	   + h^E_{ij} \superl_i\superhd\superec_j
	   + \mu \superhd\superhu.
\label{eq:2}
\ee
and
\be
W_{\Rslash} = \lambda_{ijk} \superl_i\superl_j\superuc_k
           + \lambda_{ijk}'  \superl_i\superq_j\superdc_k
	   + \lambda_{ijk}''  \superuc_i\superdc_j\superdc_k
	    + \mu_j\superl_j\superhu.
\label{eq:3}
\ee
Here $ \superq, \superl$ denote the quark and lepton SU(2) 
doublet superfields, 
$ \superuc, \superdc, \superec $ 
the corresponding SU(2) singlets and 
$ \superhu, \superhd $ the Higgs superfields whose scalar components 
give mass to up- and down-type quarks and leptons. 
In the R--parity breaking part we set $\lambda_{ijk} = 
\lambda_{ijk}''
= 0$ to avoid the unsuppressed proton decay.  

Since supersymmetry in the low-energy world is broken, the
Lagrangian of the theory is supplemented with the "soft" 
supersymmetry breaking terms: 
They are given by:
\bea
-\Lsoft
& = & \left(A^U_{(ij)} h^U_{ij} \tildeQ_i H_u \tildeuc_j
    + A^D_{(ij)} h^D_{ij} \tildeQ_i H_d \tildedc_j
    + A^E_{(ij)} h^E_{ij} \tildeL_i H_d \tildeec_j + \hc \right)  \nonumber \\
&   & \mbox{} + B\mu \left(H_d H_u + \hc \right)
    + m_{H_d}^2\vert H_d\vert^2 + m_{H_u}^2\vert H_u\vert^2 \nonumber\\
&   & \mbox{} + m_{\tildeL}^2\vert \tildeL\vert^2 + m_{\tildeec}^2\vert
\tildeec\vert^2
     + m_{\tildeQ}^2\vert \tildeQ\vert^2 + m_{\tildeuc}^2\vert \tildeuc\vert^2
     + m_{\tildedc}^2\vert \tildedc\vert^2 \nonumber \\
&   & \mbox{} + \left(\onehalf M_1 \bar{\psi}_B\psi_B + \onehalf
	M_2\bar{\psi}^a_W
	\psi^a_W + \onehalf \mgluino\bar{\psi}^a_g\psi^a_g + \hc \right)
\label{eq:4}
\eea
and
\bea
-{\Lsoft}_{\Rslash} =
      {\tilde \lambda}_{ijk} \tildeL_i\tildeL_j{\tilde u}^c_k
    + {\tilde \lambda_{ijk}}'  \tildeL_i\tildeQ_j{\tilde d}^c_k
+ {\tilde \lambda_{ijk}}'' {\tilde u}^c_i{\tilde d}^c_j{\tilde d}^c_k 
    + {\tilde\mu}^2_{2j}\tildeL_j\superhu 
    +  {\tilde\mu}^2_{1j}\tildeL_j\superhd.
\label{eq:5}
\eea
Here, the fields with tilde denote the scalar partners of the quark
and lepton fields, while the $\psi_i$ are the spin-$1 \over 2$
partners of gauge bosons.

In the present paper we concentrate  on the trilinear terms
R-parity breaking superpotential only, leaving complete treatment of 
bilinear terms for future studies.  

Since our main goal is the \rp MSSM description of the $0\nu\beta\beta$-decay,
 we have to  derive the effective Lagrangian for that process from
the superpotential. We will show  below that  
this Lagrangian depends on many free supersymmetric parameters, 
like masses or coupling constants. In order to reduce the number of 
these parameters, we apply conditions motivated by supergravity theories unification 
 for soft mass parameters at the GUT scale 
(see e.g. \cite{BBO1,KKRW,deBoer}). We shortly present our method 
of finding the low-energy spectrum of the MSSM below. 
	
We begin our procedure with running the gauge and Yukawa coupling 
constants up to the point where they unify (GUT scale).
The initial conditions for the gauge couplings are set up by the
values of $\alpha_{em}(\mz) = 1/127.89$ and $\alphas = 0.118$. 
The values of Yukawa couplings at $\mz$ are connected with the standard model
quark and lepton mass matrices in the usual way:
\bea
M_U = \frac{v \sin\beta}{\sqrt{2}} S^{U_R}Y_U^TS^{U_L \dagger}, \nonumber \\
M_D = \frac{v \cos\beta}{\sqrt{2}} S^{D_R}Y_D^TS^{D_L \dagger}, \nonumber \\
M_E = \frac{v \cos\beta}{\sqrt{2}} S^{E_R}Y_E^TS^{E_L \dagger}.
\label{eq:6}
\eea
Here, the unitary matrices S are connected with mixing among the
MSSM fields of the
matter sector induced by the electroweak symmetry breaking and 
$K = S^{U_L}S^{D_L \dagger}$ is the Kobayashi-Maskawa
matrix. In the matter sector primed mass eigenstates are related to their
unprimed gauge eigenstate counterparts.
\bea
u' &=& S^{U_L}u + S^{U_R}C{\overline{u^c}}^T, \nonumber \\
d' &=& S^{D_L}d - S^{D_R}C{\overline{d^c}}^T, \nonumber \\
e' &=& S^{E_L}e - S^{E_R}C{\overline{e^c}}^T, \nonumber \\
\nu' &=& S^{N_L}\nu.
\label{eq:7}
\eea
In above formulas, the following convention for the MSSM up an down-type
Higgs vacuum expectation values are used:
\bea
\langle H_d^0 \rangle = \frac{v_d}{\sqrt{2}}, \qquad
\langle H_u^0 \rangle = \frac{v_u}{\sqrt{2}}, \qquad
v = \sqrt{v_u^2 + v_d^2} = 246 \gev.
\label{eq:8}
\eea

During the running procedure,  the 
renormalization group equations (RGE's) change between the weak and GUT scales 
at each particles mass threshold due to decoupling of states at the scales below 
their masses. At first running, we initially set this threshold at 
$M_{SUSY} = 1 $TeV, using the SM 2-loop RG equations below, 
and the MSSM RGE's above that 
scale. When the gauge and Yukawa couplings get the GUT scale, 
we set all the soft scalar masses equal to the common scalar mass $m_0$,
the soft trilinear couplings A to common $A_0$ and take gaugino masses 
to $m_{1/2}$. In the next step one runs  everything down back to 
the $\mz$ scale. 
We want to stress at this point that we do not use the full set
of RGE's appropriate for the R--parity broken MSSM \cite{rparrge}.
We estimate that an influence of the R--parity breaking constants 
on other quantities running 
is marginal due to the smallness of $\lambda$'s.
So, we limit our attention to RGE's for MSSM with R--parity conserved
\cite{guts}).

It is well known that running of $m_{H_u}^2$ is dominated by negative 
contribution from the top Yukawa coupling, which drives this parameter to 
a negative value at some scale causing dynamical electroweak symmetry 
breaking (EWSB). Thus the RGE's improved supersymmetric potential naturally 
breaks $SU(2) \times U(1)_Y $ to $U(1)_{\rm em}$, which additionally allows 
to express some GUT-scale free parameters in terms of low-energy ones. 
The tree-level Higgs potential has the form
	
\be
V_0 = m_1^2\vert H_d^0\vert^2 + m_2^2\vert H_u^0\vert^2 +
m_3^2(H_d^0H_u^0 +
\hc) + \frac{g_1^2+g_2^2}{8}(\vert H_d^0\vert^2 - \vert
H_u^0\vert^2)^2,
\label{eq:9}
\ee
where $m_{1,2}^2\equiv m_{H_{d,u}}^2 + \mu^2, m_3^2\equiv B\mu$,
and the phases of the fields are chosen such that $m_3^2<0$.
Due to Q scale dependence of the soft Higgs parameters $V_0$ also strongly 
depends on Q. Thus minimization of $V_0$ can produce very different
vacuum expectation values 
$v_u, v_d$ depending on Q. That's why it is necessary
to minimize the full one-loop Higgs effective potential. 
The procedure leads to a set of two equations.
\bea
|\mu|^2+\frac{\mz^2}{2} & = & 
 \frac{(m^2_{H_d}+\Sigma_d)-(m^2_{H_u}+\Sigma_u)\tan^2\beta}
{\tan^2\beta -1}.
   %\nonumber 
\label{eq:10} \\
  \sin 2\beta & = & 
  \frac{-2B\mu}{(m^2_{H_u}+\Sigma_u)+(m^2_{H_d}+\Sigma_d)+2|\mu|^2}.
\label{eq:11}
\eea
$\Sigma_u, \Sigma_d$ are given e.g. in Ref. \cite{finitehiggs}.
In order to minimize the stop contribution to the finite corrections, 
we take the minimization 
scale $Q_{min}$ equal to the square root of mean
square of stop masses. 

The EWSB causes mixing among many particles. In particular, 
four gauginos mix to produce the  so-called 
neutralinos with the mass matrix $M_\chi$, which in the 
 $\psi=(\bino, \wino, \hinob, \hinot)$ basis has the form:
\begin{equation}
M_\chi =\left( 
\begin{array}{cccc}
M_1 & 0 & -m_Zc_\beta s_W & m_Zs_\beta s_W \\ 
0 & M_2 & m_Zc_\beta c_W & -m_Zs_\beta c_W \\ 
-m_Zc_\beta s_W & m_Zc_\beta c_W & 0 & -\mu  \\ 
m_Zs_\beta s_W & -m_Zs_\beta c_W & -\mu  & 0
\label{eq:12}
\end{array}
\right), 
\end{equation} 
with $c_w = \cos(\theta_W), s_w = \sin(\theta_W), 
s_\beta = \sin(\beta), c_\beta = \cos(\beta)$ 
and $M_1, M_2$ being  U(1) and SU(2) gaugino masses. 
At this point it is worth mentioning that 
the physical gluino mass is related to the renormalized 
$\overline{DR}$ running mass $m_3$. 
For very heavy quarks this dependence can read \cite{finitegluino}
\be
m_{\tilde{g}}^{\rm pole} \simeq m_3 \left[ 1 + {\alpha_3 \over 4 \pi} 
  \Big( 15 + 12 I(r) \Big) \right],
\label{eq:13}
\ee
where $m_3$ and $\alpha_3$ are taken at $m_3$ and the loop function 
$I(r) = {1 \over 2} \ln r + {1 \over 2} (r-1)^2 \ln(1-r^{-1}) 
 + {1 \over 2} r -1$ for $r \geq 1$ with
$r = m_{\tilde{q}}^2 / m_3^2$.

Mixing in the gaugino sector effects with four physical 
neutralinos $\chi_{i}$.
\begin{equation}
\chi _{i}=\sum_{j=1}^{4}{\cal N}_{ij}\psi _{j}, \qquad (i = 1, 2, 3, 4).
\label{eq:14}
\end{equation}
The matrix ${\cal N}$, diagonalizing the
 matrix $M_{\chi }$
is real and orthogonal. Thus, with real ${\cal N}_{ij}$, neutralino
 masses are either positive or negative. 
If necessary, a negative mass can always be made positive by 
a redefinition of the relevant mixing 
coefficients ${\cal N}_{ij} \to i{\cal N}_{ij}$. 

Similar mixing appears in the slepton and squark sector. 
The SUSY
analog of mixing (\ref{eq:7}) in the SM sector is described  as follows:
\bea
{\tilde u}' &=& \Gamma^U\vect{S^{U_L}{\tilde u}}{S^{U_R}{\tilde u}^{c*}},
\nonumber \\ 
{\tilde d}' &=& \Gamma^D\vect{S^{D_L}{\tilde d}}{-S^{D_R}{\tilde d}^{c*}},
\nonumber \\
{\tilde e}' &=& \Gamma^E\vect{S^{E_L}{\tilde e}}{-S^{E_R}{\tilde e}^{c*}}.
\nonumber \\
{\tilde \nu}' &=& \Gamma^N S^{E_L} {\tilde \nu}, 
\label{eq:15}
\eea
The $6 \times 6$ squared mass matrices for the squarks and sleptons have a
much more complicated form and involve parameters from both the
supersymmetry breaking and conserving Lagrangians.
They can be written in the following form:
\bea
\label{eq:16}
M_{\tilde u}^2 = \Gamma^U
\left(
\begin{array}{c|c}
S^{U_L}m_{\tilde Q}^2 S^{U_L \dagger} + M_U^2 &
-\mu M_U \cot\beta   \\
+ \frac{\mz^2}{6}(3 - 4 \swsq) \cos2\beta &
- \frac{v \sin\beta}{\sqrt{2}}S^{U_L}A^{U*}S^{U_R \dagger} 
\\
\hline
-\mu^* M_U \cot\beta  &
S^{U_R}{m_{{\tilde u}^c}^2}^T S^{U_R \dagger} + M_U^2 \\
- \frac{v \sin\beta}{\sqrt{2}}S^{U_R}A^{U^T}S^{U_L \dagger} &
+ \frac{2\mz^2}{3} \swsq\cos2\beta  
\end{array}
\right)\Gamma^{U \dagger},
\eea
\bea
\label{eq:17}
M_{\tilde d}^2 = \Gamma^D
\left(
\begin{array}{c|c}
S^{D_L}m_{\tilde Q}^2 S^{D_L \dagger} + M_D^2 &
-\mu M_D \tan\beta   \\
- \frac{\mz^2}{6}(3 - 2 \swsq) \cos2\beta &
- \frac{v \cos\beta}{\sqrt{2}}S^{D_L}A^{D*}S^{D_R \dagger}  \\
\hline
-\mu^* M_D \tan\beta &
S^{D_R}{m_{{\tilde d}^c}^2}^T S^{D_R \dagger} + M_D^2  \\
- \frac{v \cos\beta}{\sqrt{2}}S^{D_R}A^{D^T}S^{D_L \dagger} &
- \frac{\mz^2}{3} \swsq\cos2\beta
\end{array}
\right)  \Gamma^{D \dagger}, 
\eea
\bea
\label{eq:18}
M_{\tilde e}^2 = \Gamma^E 
\left(
\begin{array}{c|c}
S^{E_L}m_{\tilde l}^2 S^{E_L \dagger} + M_E^2 &
-\mu M_E \tan\beta   \\
- \frac{\mz^2}{2}(1 - 2 \swsq) \cos2\beta &
- \frac{v \cos\beta}{\sqrt{2}}S^{E_L}A^{E*}S^{E_R \dagger} \\
\hline
-\mu^* M_E \tan\beta  &
S^{E_R}{m_{{\tilde e}^c}^2}^T S^{E_R \dagger} + M_E^2  \\
- \frac{v \cos\beta}{\sqrt{2}}S^{E_R}A^{E^T}S^{E_L \dagger} &
- \mz^2 \swsq\cos2\beta
\end{array}
\right) 
  \Gamma^{E \dagger}. 
\eea

When the physical masses of 
scalar particles and tree level values of $\mu$ and $B\mu$ are found, 
one can set up thresholds
 for RGE's of the gauge and Yukawa couplings and repeat the 
 procedure by running everything up to the 
 GUT scale. Setting the unification conditions 
 for soft parameters again, it is possible to run everything back 
 to the $\mz$ scale. At that point we repeat calculation of mass
  eigenstates, run everything up to $Q_{\rm min}$ 
 and minimize the one-loop corrected Higgs potential 
 mentioned above. Repeating such a procedure 
 leads to fixed  values of $\mu$ and $B\mu$. At the end,  
 we check for conditions necessary for existence  of
a stable scalar potential minimum:
\bea
(\mu B)^2 &>& \left( \left|\mu\right|^2 + \mHu^2 \right)
\left( \left|\mu\right|^2 + \mHd^2 \right), 
\nonumber \\
2B\mu &<& 2 \left|\mu\right|^2 + \mHu^2 + \mHd^2.
\label{eq:19}
\eea

%%%%%%%%%%%%%%%%%      CONSTRAINTS       %%%%%%%%%%%%%%%%%%%%%%%%%%%%%%%
\subsection{Constraints}
%%%%%%%%%%%%%%%%%%%%%%%%%%%%%%%%%%%%%%%%%%%%%%%%%%%%%%%%%%%%%%%%%%%%%%%%

Having obtained the low energy spectrum of the model, we are at a good 
point to test 
our scenario using constraints due to rare Flavor Changing Neutral
Currents processes. It is well known that FCNC processes may serve as a
strong constraint on the supersymmetric scenarios.  Strong experimental
suppression of FCNC transition puts upper bounds on various entries of the
sfermion mass matrices at low energy. Many analyses of such constraints were
performed for the SUGRA MSSM \cite{bsgmssm}. Some of them are also  
available for the Gauge Mediated Supersymmetry Breaking (GMSB)
\cite{bsggmsb}. Without going into details of calculations, we
summarize the basic ingredients of the applied procedure. 

In our considerations we concentrate on limits on the low-energy spectrum
coming from the $B \to X_s \gamma$ decay. This process is described by an
effective Hamiltonian of the form \cite{bsgmssm,bsg_pokorscy}:
\bea
H_{\rm eff} = - \frac{4G_F}{\sqrt{2}}K^*_{ts}K_{tb}\sum_{i =
1}^8C_i(\mu)P_i(\mu), 
\label{e:20}
\eea
where $K$ is the Kobayashi-Maskawa matrix, $P_i$ are the relevant operators,
and $C_i(\mu)$ are their Wilson coefficients. The coefficients 
$C_7$ and $C_8$ relevant in the subsequent
analysis   get contributions from both the SM
and the MSSM interactions. The leading order and next-to-leading order SM
contributions  are discussed in \cite{bsgnlo}. For the MSSM case, only
leading order contributions to those coefficients are available
\cite{bsgmssm,bsg_pokorscy}
\footnote{NLO QCD corrections for SUSY scenario with  charginos and one of the stops 
lighter than other sparticles can be found in \cite{bsg_mssm_nlo}.}. 

The constraints on the low-energy spectrum of our model are connected
with present constraints on the $R_7$ parameter which measures the extra
(MSSM) contributions to the $B \to X_s \gamma$ decay. Its definition reads:
\bea
R_7 \equiv 1 + \frac{C_7^{(0)extra}(\mw)}{C_7^{(0)SM}(\mw)},
\label{eq:21}
\eea
where index $(0)$ stands for the leading order (LO) Wilson coefficients, 
and superscript $extra$ for the SUSY (namely charged Higgs, chargino, 
neutralino and gluino) contributions.
 
Limits on the allowed values of $R_7$ are extracted from the present 
experimental limits on the branching ratio BR($B \to X_s \gamma$)
recently measured by CLEO collaboration
\cite{bsg_mssm_nlo}:
BR$(B \to  X_s \gamma) = (2.50 \pm 0.47_{stat} \pm 
0.39_{syst})\times 10^{-4}$. 
The same process measured at ALEPH collaboration
results with 
BR$(B \to  X_s \gamma) = (3.11 \pm 0.80_{stat} \pm 
0.72_{syst})\times 10^{-4}$
\cite{ALEPH}.
Using the theoretical dependence of the theoretical expectations of
BR($B \to X_s \gamma$) on $R_7$, comparison with the experimental data
\cite{bsg_pokorscy} and taking into account both theoretical and experimental
errors, we end up with the following estimation of the $R_7$ range :
\bea
-6.6 < R_7 < -4.4 \,\,\,\,{\mbox {or}}\,\,\,\, 0.0 < R_7 < 1.3.
\label{eq:22}
\eea

In order to find numerical values of $R_7$ given in (\ref{eq:21}), we use the
expressions for  $C_7^{(0)MSSM}$ and $C_7^{(0)SM}$ listed  in e.g.
\cite{bsgmssm}.
Then we exclude points which do not lie in the allowed region (\ref{eq:22}).
In addition to $B \to  X_s \gamma$ constraints, the free parameter space
is limited by the conditions of proper electroweak symmetry breaking 
(\ref{eq:19}) and the condition of positive (mass)$^2$ of mass eigenstates.
In the last section we present the resulting constraints in $\tan\beta - m_{0}$
and $\tan\beta - m_{1/2}$ planes for $\mu$ of different signs. Moreover, 
the detailed analysis of $B \to  X_s \gamma$ constraints is shown.

%%%%%%%%%%%%%%%%%%%%%%%%%%%%%%%%%%%%%%%%%%%%%%%%%%%%%%%%%%%%%%%%%
\subsection{R-parity violating neutrinoless double beta decay}
%%%%%%%%%%%%%%%%%%%%%%%%%%%%%%%%%%%%%%%%%%%%%%%%%%%%%%%%%%%%%%%%%

In this Subsection the main formulae relevant for R-parity violating
$0\nu\beta\beta$-decay are presented. A detailed derivation of the
effective \rp quark-level Lagrangian for 
the $0\nu\beta\beta$-decay and its hadronization additionaly to derivation of the
corresponding half-life time of this process can be found elsewhere
Refs. \cite{HKS96b,WKS97,FKSS97,FKS98a}. 

The trilinear R-parity violating lepton part of the 
interaction Lagrangian of the MSSM takes the form \cite{HKS96b}:
\begin{eqnarray}
  {\cal L}_{\lambda'_{111}}=
  -\lambda'_{111}[({\bar{u}_{L}},{\bar{d}_{L}})
     \left(
         \begin {array}{c}
           e_{R}^{c}\\-\nu_{R}^{c}
         \end{array}
     \right)
  \tilde{d}_{R}^{\ast}+({\bar{e_{L}}},{\bar{\nu_{L}}}) d_{R}
     \left(
         \begin {array}{c}
           \tilde{u}^{\ast}_{L} \\ -\tilde{d}^{\ast}_{L}
         \end{array}
     \right)
 + \nonumber \\
  ({\bar{u_{L}}},{\bar{d_{L}}}) d_{R}
     \left(
         \begin {array}{c}
           \tilde{e}^{\ast}_{L} \\ -\tilde{\nu}^{\ast}_{L}
         \end{array}
     \right)
 + \hc].
\label{eq:23}
  \end{eqnarray}
Together with the R-parity conserving 
MSSM Lagrangian describing interactions among gluinos, neutralinos, 
fermions and sfermions 
\cite{haberkane} 
after  integration out of heavy degrees of freedom and carrying 
 a Fierz transformation, one can obtain finally 
the \rp SUSY induced 
quark-lepton interaction for $0\nu\beta\beta$-decay.
\begin{eqnarray}
 {\cal L}^{\Delta L_e =2}_{\rm eff}\ =
\frac{G_F^2}{2 m_{_p}}~ \bar e (1 + \gamma_5) e^{\bf c} 
\left[\eta_{PS}~J_{PS}J_{PS} 
- \frac{1}{4} \eta_T\   J_T^{\mu\nu} J_{T \mu\nu} \right].
\label{eq:24}
\end{eqnarray}
The color singlet hadronic currents in Eq. (\ref{eq:24}) are 
$J_{PS} =   {\bar u}^{\alpha} \gamma_5 d_{\alpha} + 
{\bar u}^{\alpha} d_{\alpha}$, 
$J_T^{\mu \nu} = {\bar u}^{\alpha} 
\sigma^{\mu \nu} (1 + \gamma_5) d_{\alpha}$, 
where $\alpha$ is a color index and 
$\sigma^{\mu \nu} = (i/2)[\gamma^\mu , \gamma^\nu ]$.
The effective lepton-number violating parameters $\eta_{PS}$
and $\eta_{T}$  in Eq.\ (\ref{eq:24}) accumulate fundamental parameters of 
the \rp MSSM:
\begin{eqnarray}
\label{eq:25}
\eta_{PS} &=&  \eta_{\chi\tilde e} + \eta_{\chi\tilde f} +
\eta_{\chi} + \eta_{\tilde g} + 7 \eta_{\tilde g}^{\prime}, \\
\label{eta}
\eta_{T} &=& \eta_{\chi} - \eta_{\chi\tilde f} + \eta_{\tilde g}
- \eta_{\tilde g}^{\prime},
\label{eq:26}
\end{eqnarray}
with
\begin{eqnarray}
\eta_{\tilde g} &=& \frac{\pi \alpha_s}{6}
\frac{\lambda^{'2}_{111}}{G_F^2 m_{\tilde d_R}^4} \frac{m_p}{m_{\tilde 
g}}\left[
1 + \left(\frac{m_{\tilde d_R}}{m_{\tilde u_L}}\right)^4\right],
\nonumber\\
%%%%
\eta_{\chi} &=& \frac{ \pi \alpha_2}{2}
\frac{\lambda^{'2}_{111}}{G_F^2 m_{\tilde d_R}^4}
\sum_{i=1}^{4}\frac{m_p}{m_{\chi_i}}
\left[
\epsilon_{R i}^2(d) + \epsilon_{L i}^2(u)
\left(\frac{m_{\tilde d_R}}{m_{\tilde u_L}}\right)^4\right],
\nonumber \\
%%%%
\eta_{\chi \tilde e} &=& 2 \pi \alpha_2
\frac{\lambda^{'2}_{111}}{G_F^2 m_{\tilde d_R}^4}
\left(\frac{m_{\tilde d_R}}{m_{\tilde e_L}}\right)^4
\sum_{i=1}^{4}\epsilon_{L i}^2(e)\frac{m_p}{m_{\chi_i}},
\nonumber \\
%%%%
\eta'_{\tilde g} &=& \frac{\pi \alpha_s}{12}
\frac{\lambda^{'2}_{111}}{G_F^2 m_{\tilde d_R}^4}
\frac{m_p}{m_{\tilde g}}
\left(\frac{m_{\tilde d_R}}{m_{\tilde u_L}}\right)^2,
\nonumber \\
%%%%
\eta_{\chi \tilde f} &=& \frac{\pi \alpha_2 }{2}
\frac{\lambda^{'2}_{111}}{G_F^2 m_{\tilde d_R}^4}
\left(\frac{m_{\tilde d_R}}{m_{\tilde e_L}}\right)^2
\sum_{i=1}^{4}\frac{m_p}{m_{\chi_i}}
\left[\epsilon_{R i}(d) \epsilon_{L i}(e)  + \right.
\nonumber \\
&+& \left.\epsilon_{L i}(u) \epsilon_{R i}(d)
\left(\frac{m_{\tilde e_L}}{m_{\tilde u_L}}\right)^2
+ \epsilon_{L i}(u) \epsilon_{L i}(e)
\left(\frac{m_{\tilde d_R}}{m_{\tilde u_L}}\right)^2
\right].
%%%%
\label{eq:27}
\end{eqnarray}
In Eq. (\ref{eq:27}) $G_F$ is the Fermi constant, 
$m_p$ is the proton mass,
$\alpha_2 = g_2^2/(4\pi )$ and $\alpha_s = g^2_3/(4\pi )$ are 
 $\rm SU(2)_L$ and $\rm SU(3)_c$ gauge coupling constants respectively.
$m_{{\tilde u}_L}$, $m_{{\tilde d}_R}$, 
$m_{\tilde g}$ and $m_{\chi_i}$ are masses of the u-squark, d-squark,
gluino and neutralinos. We note that
the matrix ${\cal N}_{ij}$ (see Eq. (\ref{eq:14})) rotates
the $4\times 4$ neutralino mass matrix in Eq. (\ref{eq:12}) to 
obtain the diagonal
form $Diag [m_{\chi }]$. We used the neutralino couplings 
in the form proposed in Ref. \cite{haberkane}:
\begin{eqnarray}
\epsilon_{L_i}(\phi ) & = & -T_3 (\phi ) {\cal N}_{i2} + 
tan \theta_W [T_3 (\phi ) - Q (\phi ) ] {\cal N}_{i1}, 
\nonumber \\
\epsilon_{R_i}(\phi ) & = & Q (\phi ) tan \theta_W {\cal N}_{i1}.
\label{eq:28}
\end{eqnarray}

The effective $\Delta L_e = 2$ Lagrangian (\ref{eq:24})
contains  contributions from both
gluino (SUSY parameters $\eta_{\tilde g}$ and $\eta'_{\tilde g}$) 
and neutralino (SUSY parameters $\eta_{\chi}$, $\eta_{\chi \tilde f}$ and
$\eta_{\chi \tilde e}$) exchanges. 
The relevant Feynman diagrams associated with 
gluino ${\tilde g}$ and neutralino $\chi $ contributions 
to the $0\nu\beta\beta$-decay have been discussed  in Refs. 
\cite{FS98,HKS96b}.

The \rp MSSM model  gives the underlying transition of a 
down-quarks to an up-quarks
($dd \rightarrow uu + 2 e^-$) only, 
and results in  transformation of neutron into a proton. 
Till now three possibilities of hadronization have been considered.
The most natural way is to incorporate the quarks 
in the nucleons which is the  well-known two-nucleon mode 
\cite{Moh86,Ver87,HKS96b}.
But the intermediate SUSY partners are very heavy particles. 
Therefore in the two-nucleon mode the two decaying neutrons must come
very close to each other, which is suppressed by the nucleon repulsion.
Another possibility is to incorporate quarks undergoing the
\rp SUSY transition  not in nucleons but in virtual pions 
\cite{FS98,FKSS97,FKS98b}.
If only one of the initial quark is placed in an intermediate pion, we end up
with the one pion-exchange mode \cite{FKS98b}. If all quarks are placed in two
intermediate pions, one obtains the two-pion exchange mode, which 
 dominates for the $0\nu\beta\beta$-decay of $^{76}Ge$ \cite{FKSS97,FKS98b}.
The possibilities of incorporating quarks in heavier mesons 
have not been considered
so far. However, this way of hadronization is expected to 
be suppressed due to
heavier masses of exchange particles.

The half-life for the neutrinoless double beta decay regarding the  above three
possibilities of hadronization of the quarks can be written in the
form  \cite{FS98,FKSS97,FKS98b}
\begin{equation}
\big[ T_{1/2}^{0\nu}(0^+ \rightarrow 0^+) \big]^{-1}~=~
G_{01} \left | \eta_{T}~ {\cal M}_{\tilde q}^{2N}
 ~+ ~ \Big(\eta_{PS} ~-~ \eta_{T}\Big)~ {\cal M}_{\tilde f}^{2N} ~+ ~\frac{3}{8}
 \left(\eta_{T}~ + ~\frac{5}{8} \eta_{PS}\right)~ {\cal M}^{\pi N} \right |^2.
\label{eq:29}
\end{equation}
where 
\begin{eqnarray}
{\cal M}^{2N}_{\tilde q} &=&  c_A \Big[
\alpha^{(0)}_{V-\tilde{q}} {\cal M}_{F N} + 
\alpha^{(0)}_{A-\tilde{q}} {\cal M}_{GT N} +
\alpha^{(1)}_{V-\tilde{q}} {\cal M}_{F'} + 
\alpha^{(1)}_{A-\tilde{q}} {\cal M}_{GT'} +
\alpha_{T-\tilde{q}} {\cal M}_{T'} \Big]\,, 
\label{eq:30} \\
{\cal M}^{2N}_{\tilde f} &=& c_A \Big[
\alpha^{(0)}_{V-\tilde{f}} {\cal M}_{F N} + 
\alpha^{(0)}_{A-\tilde{f}} {\cal M}_{GT N} +
\alpha^{(1)}_{V-\tilde{f}} {\cal M}_{F'} + 
\alpha^{(1)}_{A-\tilde{f}} {\cal M}_{GT'} +
\alpha_{T-\tilde{f}} {\cal M}_{T'} \Big]\,, 
\label{eq:31} \\
{\cal M}^{\pi N} &=& c_A \Big[
 \frac{4}{3}\alpha^{1\pi}\left(M_{GT-1\pi} + M_{T-1\pi} \right)
      +
      \alpha^{2\pi}\left(M_{GT-2\pi} + M_{T-2\pi} \right)\Big]\, 
\label{eq:32}
\end{eqnarray}
with $c_{_{A}} = m^2_{_{A}}/(m_p m_e)$.
Here $G_{01}$ is the standard phase space factor 
(see Ref. \cite{doi83,PSVF96})
and $m_A = 850$ MeV is the nucleon form factor cut-off (for all nucleon
form factors the dipole shape with the same cut-off is considered).
The nucleon structure coefficients $\alpha 's$ entering 
the nuclear matrix elements of the two-nucleon mode calculated 
within the non-relativistic quark model (NR) and the bag model (BM) 
are given in Table 3 of Ref. \cite{FS98}.
The structure coefficient of the one-pion  $\alpha^{1\pi}$
and two-pion mode $\alpha^{2\pi}$ are \cite{FKSS97,FKS98b}:
$\alpha^{1\pi} = -0.044$ and $\alpha^{2\pi} = 0.20 $. 
The partial nuclear matrix elements of the
$R_p \hspace{-1em}/\;\:$  SUSY mechanism for the $0\nu\beta\beta$-decay 
appearing in Eqs. (\ref{eq:30})-(\ref{eq:32})
are:
\begin{eqnarray}
{\cal M}_I &=&
\langle 0^+_f|~\sum_{i\neq j} ~\tau_i^+ \tau_j^+ ~
\frac{R_0}{r_{ij}}
~F_I(x)
~| 0^+_i \rangle , \nonumber \\ 
{\cal M}_J &=&
\langle 0^+_f|~\sum_{i\neq j} ~\tau_i^+ \tau_j^+ ~
\frac{R_0}{r_{ij}}
~F_J(x)
~{\bf{\sigma}}_i\cdot{\bf{\sigma}}_j, 
~| 0^+_i \rangle , \nonumber \\ 
{\cal M}_K &=&
\langle 0^+_f|~\sum_{i\neq j} ~\tau_i^+ \tau_j^+ ~
\frac{R_0}{r_{ij}}
~F_K(x)
~{\bf{S}}_{ij}
~| 0^+_i \rangle , 
\label{eq:33}
\end{eqnarray}
where $I = FN,~ F'$, $J = GTN,~ GT', ~ GT-k\pi$ and 
$K = T',~ T-k\pi$ ($k=1,2$).
The structure functions $F_{I,J,K}$ can be expressed as:
\begin{eqnarray}
&&F_{FN}(x) = F_{GTN}(x) = 
\frac{x_{_{A}}}{48}(3 + 3x_{_{A}} + x^2_{_{A}}) e^{-x_{_{A}}},\nonumber \\
&&F_{F'}(x) = F_{GT'}(x) = 
\frac{x_{_{A}}}{48}
(3 + 3x_{_{A}} - x^2_{_{A}}) e^{-x_{_{A}}},\nonumber \\ 
&&F_{T'}(x) = \frac{x^2_{_{A}}}{48}e^{-x_{_{A}}},
\nonumber \\
&&F_{GT-1\pi}(x_\pi ) = e^{-x_{{\pi}}}, \nonumber \\ 
&&F_{T-1\pi}(x_\pi )  = (3 + 3x_{{\pi}} + x^2_{{\pi}} )
\frac{e^{-x_{{\pi}}}}{x^2_{{\pi}}}, \nonumber \\
&&F_{GT-2\pi}(x_\pi ) = (x_{{\pi}} - 2) ~e^{-x_{{\pi}}}, \nonumber \\ 
&&F_{T-2\pi}(x_\pi )  = (x_{{\pi}}+1) ~e^{-x_{{\pi}}}.
\label{eq:34}
\end{eqnarray}
and we used the notations: 
${\bf{\hat{r}}}_{ij} =$ $(\bf{r}_i - \bf{r}_j)/
|\bf{r}_i - \bf{r}_j|$, $r_{ij}=|\bf{r}_i - \bf{r}_j|$,
$x_{_{A}} = m_{_{A}} r_{ij}$, $x_{\pi} = m_{{\pi}} r_{ij}$ and 
${\bf{S}}_{ij}= 3 {\bf{\sigma}}_i \cdot {\bf{\hat{r}}}_{ij} ~
 {\bf{\sigma}}_j \cdot {\bf{\hat{r}}}_{ij} -
{\bf{\sigma}}_i \cdot {\bf{\sigma}}_j$.   
Here  $\bf{r}_i$ is the coordinate of the i-th nucleon
and $R = r_0 A^{1/3}$ 
stands for the mean nuclear radius with $r_0 = 1.1$ fm.

In order to deduce constraints on the lepton number violating parameters
$\eta_{PS}$ and $\eta_{T}$ from the non-observability of 
$0\nu\beta\beta$-decay 
it is necessary to evaluate the nuclear matrix elements
of two-nucleon (${\cal M}^{2N}_{\tilde q}$, ${\cal M}^{2N}_{\tilde f}$)
and pion-exchange  (${\cal M}^{\pi N}$) modes 
within an appropriate nuclear model.

%%%%%%%%%%%%%%%%%%%%%%%%%%%%%%%%%%%%%%%%%%%%%%%%
\section{Nuclear Model}
%%%%%%%%%%%%%%%%%%%%%%%%%%%%%%%%%%%%%%%%%%%%%%%%

We calculate the nuclear matrix elements within
the proton-neutron renormalized
Quasiparticle Random Phase Approximation (pn-RQRPA) \cite{TS95,SSF96},
which is an extension of the pn-QRPA \cite{VZ86,CFT87} by incorporating 
the Pauli exclusion principle for the fermion pairs. The advantage of the
pn-RQRPA over the usual pn-QRPA is no collapsing 
RQRPA solution as one  increases  the strength of the 
particle-particle interaction within its physical values. Thus 
the results obtained by the pn-RQRPA method are more reliable, but
the pn-RQRPA method requires coupled non-linear 
equation solutions, instead of the usual eigenvalue problem in  the 
pn-QRPA formalism. 

The pn-RQRPA method is suitable to deal with the nuclear
structure aspects of beta transitions of  open shell systems and
allows to perform calculations in realistic 
large model spaces, which are unaccessible for the shell model
calculations.

The pn-RQRPA formalism consists of two main steps: (i) The Bogoliubov
transformation smears out the nuclear Fermi surface over a relatively
large number of orbitals and (ii) the equation of motion in the quasiparticle
basis determines then the excited states. 

If only proton--proton and neutron--neutron pairing is considered  
the particle ($c^{+}_{\tau m_\tau}$ and
$c^{}_{\tau m_\tau}$, $\tau = p,n$) and quasiparticle 
($a^{+}_{\tau m_\tau}$ and $a^{}_{\tau m_\tau}$, $\tau = p,n$) 
creation and annihilation operators for the 
spherical shell model states  are related to each other by the 
Bogoliubov-Valatin transformation: 
\begin{equation}  
\left( \matrix{ c^{+}_{\tau m_{\tau} } \cr
 {\tilde{c}}_{\tau  m_{\tau} } 
}\right) = \left( \matrix{ 
u_{\tau} & -v_{\tau} \cr 
v_{\tau} & u_{\tau} 
}\right)
\left( \matrix{ a^{+}_{\tau m_{\tau}} \cr
{\tilde{a}}_{\tau m_{\tau}} 
}\right),
\label{eq:35}
\end{equation} 
where the tilde indicates the  time-reversal operation
$\tilde{a}_{\tau {m}_{\tau}}$ = 
$(-1)^{j_{\tau} - m_{\tau}}a^{}_{\tau -m_{\tau}}$. 
The occupation amplitudes $u$ and $v$ and the single quasiparticle
energies $E_\tau$ are obtained by solving the BCS equation. 
Then one gets a nuclear
Hamiltonian in quasiparticle representation
\begin{equation}
H = \sum_{\tau m_\tau} E_\tau a^+_{\tau m_\tau}a_{\tau m_\tau} + 
H_{22} + H_{40} + H_{04} + H_{31} + H_{13},
\label{eq:36}
\end{equation}
where $H_{ij}$ is the normal ordered part of the residual interaction with 
$i$ creation and $j$ annihilation operators (see e.g. Ref. \cite{ROWE}).

In the framework of the pn-RQRPA
the $m^{th}$ excited state with the angular momentum $J$ and the projection $M$ 
is created by a phonon-operator $Q^{m\dagger}_{JM^\pi}$
\begin{equation}
|m,JM\rangle = Q^{m\dagger}_{JM^\pi}|0^+_{RPA}\rangle
 \qquad \mbox{with} \qquad
Q^{m}_{JM^\pi}|0^+_{RPA}\rangle=0.
\label{eq:37}
\end{equation}
Here $|0^+_{RPA}\rangle$ is the ground state of the initial or 
the final nucleus and
the phonon-operator $Q^{m\dagger}_{JM^\pi}$  is defined by the ansatz:
\begin{equation}
Q^{m\dagger}_{JM^\pi}=\sum_{pn}
  X^m_{(pn, J^\pi)} A^\dagger(pn, JM)
+ Y^m_{(pn, J^\pi)}\tilde{A}(pn, JM).
\label{eq:38}
\end{equation}
 $ A^\dagger(pn, JM)$ 
($ A^{}(pn, JM)$) 
is the two quasi-particle creation (annihilation) 
operator coupled to the good
angular momentum $J$ with projection $M$, namely
\begin{equation}
A^\dagger(pn, JM) 
= \sum^{}_{m_p , m_n }
C^{J M}_{j_p m_p j_n m_n } a^\dagger_{p m_p} a^\dagger_{n m_n}.
\label{eq:39}
\end{equation}

In the pn-RQRPA the commutator of two-bifermion operators
is replaced by its expectation value in
the correlated QRPA ground state $|0^+_{QRPA}>$
(renormalized quasiboson approximation). Therefore we have
\begin{eqnarray}
&& \big<0^+_{RPA}\big|\big
[A^{} (pn, JM), A^+(p'n', JM)\big]
\big|0^+_{RPA}\big> = \delta_{pp'}\delta_{nn'}\times
\nonumber \\ 
\lefteqn{\underbrace{
\Big\{1
\,-\,\frac{1}{\hat{\jmath}_{l}}
<0^+_{RPA}|[a^+_{p}{\tilde{a}}_{p}]_{00}|0^+_{RPA}>
\,-\,\frac{1}{\hat{\jmath}_{k}}
<0^+_{RPA}|[a^+_{n}{\tilde{a}}_{n}]_{00}|0^+_{RPA}>
\Big\}
}_{
=:\displaystyle {\cal D}_{pn, J^\pi}
},} &&
\nonumber \\
\label{eq:40}
\end{eqnarray}
with  $\hat{\jmath}_p=\sqrt{2j_p+1}$.
Replacing $|0^+_{RPA}>$ in Eq. (\ref{eq:40}) by the uncorrelated
BCS ground state leads to  the usual quasiboson approximation 
( ${\cal D}_{pn} = {\bf 1}$), which violates the Pauli 
exclusion principle by neglecting the 
terms coming from the commutation
rules of the quasi-particles.  
From Eqs. (\ref{eq:37}) 
one can derive the RQRPA equation
\begin{equation}
  \underbrace{{\cal D}^{-1/2}\left(
    \begin{array}{cc}
       \cal A &\cal B\\
       \cal -B &\cal -A
    \end{array}
    \right)
    {\cal D}^{-1/2}}
    _{ \textstyle \overline{\cal A},\overline{\cal B}}
    \;
    \underbrace{{\cal D}^{1/2}\left(
    \begin{array}{c}
       X^m\\
       Y^m
    \end{array}
    \right)}_{\textstyle \overline{X}^m, \overline{Y}^m}
    = \Omega^m_{J^\pi}
    \underbrace{{\cal D}^{1/2}
    \left(
    \begin{array}{c}
       X^m\\
       Y^m
    \end{array}
    \right)}_{\textstyle\overline{X}^m, \overline{Y}^m}\, .
\label{eq:41}
\end{equation}
The matrices $\overline{\cal{A}}$ and
$\overline{\cal{B}}$  are given as follows:
\begin{eqnarray}
\label{eq:42}
{\overline{{\cal A}}}^{J^\pi}_{pn,p'n'} &=& D^{-1/2}_{pn, J^\pi}
\left\langle 0^+_{RPA}\right|\left[
A (pn, JM)
,\left[H,A^{\dagger} (p'n', JM)
\right]\right]\left|0^+_{RPA}\right\rangle\, D^{-1/2}_{p'n', J^\pi}
\nonumber \\
&=& {E_p+E_n}\delta_{pp'}\delta_{nn'} -  2 [~
G(pn, p'n'; J) (u_p u_n u_{p'} u_{n'}+u_p u_n u_{p'} u_{n'}) -\nonumber \\
&& ~~~~~~~~~~~~~
F(pn, p'n'; J) (u_p v_n u_{p'} v_{n'}+v_p u_n v_{p'} u_{n'})
~] ~{D^{1/2}_{pn, J^\pi}} {D^{1/2}_{p'n', J^\pi}}, \\
\label{eq:43}
{\overline{\cal B}}^{J^\pi}_{pn, p'n'} &=& D^{-1/2}_{pn, J^\pi}
\left\langle 0^+_{RPA}\right|\left[
A (pn, JM),\left[
H,{\tilde{A}} (p'n', JM)\right]\right]\left|0^+_{RPA}\right\rangle\,
D^{-1/2}_{p'n', J^\pi} \nonumber \\
&=& 2 {D^{1/2}_{pn, J^\pi}} {D^{1/2}_{p'n', J^\pi}}~ [~
G(pn, p'n'; J) (u_p u_n v_{p'} v_{n'}+v_p v_n u_{p'} u_{n'}) -\nonumber \\
&& ~~~~~~~~~~~~~
F(pn, p'n'; J) (u_p v_n v_{p'} u_{n'}+v_p u_n u_{p'} v_{n'}) ~].
\end{eqnarray}
Here, $G(pn, p'n', J)$ and  $F(pn, p'n', J)$ are the 
particle-particle and particle-hole interaction matrix elements,
respectively \cite{ROWE}.
The coefficients 
${\cal D}_{pn, J^\pi}$ 
are determined by solving numerically the system of 
equations\cite{SC98,FS98}:
\begin{eqnarray}
{\cal D}_{pn, J^\pi} &=& 1-\frac{1}{2 j_p + 1} 
\sum_{n' \atop J'^{\pi'} m}
{\cal D}_{pn', J'^{\pi'}}\hat{J}'^2 \big|
{\overline{Y}}^m_{(pn', J'^{\pi'})}  \big|^2
\nonumber \\
&& ~~~~~-\frac{1}{2j_n +1}\sum_{p' \atop J'^{\pi'} m}
{\cal D}_{p'n, J'^{\pi'}}
\hat{J}'^2 \big |{\overline{Y}}^m_{(p'n, J'^{\pi'})}
  \big|^2 .
\label{eq:44}
\end{eqnarray}
The selfconsistent scheme of the calculation of the forward-
(backward-) going free variational amplitudes   
${\overline{X}}^m_{}$ (${\overline{Y}}^m_{}$), 
the excited energies $\Omega^m_{J^\pi}$ related to the ground state 
and the coefficients ${\cal D}_{pn, J^\pi}$ 
is a double iterative problem which requires the solution of 
coupled Eqs. (\ref{eq:41}) and (\ref{eq:44}).

Numerical treatment of the 
matrix elements (\ref{eq:30})-(\ref{eq:33})
within the pn-RQRPA needs transformation of
them to  relative coordinates.
After some tedious algebra one can obtain
\begin{eqnarray}
\label{eq:45}
M^I_{type} =
\sum_{{p n p' n' } \atop {J^{\pi}
m_i m_f {\cal J}  }}
~(-)^{j_{n}+j_{p'}+J+{\cal J}}(2{\cal J}+1)
\left\{
\matrix{
j_p &j_n &J \cr
j_{n'}&j_{p'}&{\cal J}}
\right\}\times~~~~~~~~\nonumber \\
<p(1), p'(2);{\cal J}|f(r_{12})\tau_1^+ \tau_2^+ 
{\cal O}_{type}^I (12)
f(r_{12})|n(1) ,n'(2);{\cal J}>\times ~~~~
\nonumber \\
< 0_f^+ \parallel
\widetilde{[c^+_{p'}{\tilde{c}}_{n'}]_J} \parallel J^\pi m_f>
<J^\pi m_f|J^\pi m_i>
<J^\pi m_i \parallel [c^+_{p}{\tilde{c}}_{n}]_J \parallel
0^+_i >.
\end{eqnarray}
In Eq. (\ref{eq:45}) ${\cal O}_{type}^I (12)$ 
represents the coordinate and spin dependent part of the nuclear
two body transition operator for the $0\nu\beta\beta$-decay
and can be expressed in the form:
\begin{equation}
{\cal O}_{type}^I (12) = 
H_{type-F}^{I}(r_{12})   + 
H_{type-GT}^{I}(r_{12}) {\bf \sigma}_{12} +
H_{type-T}^{I}(r_{12}) {\bf S}_{12}.
\label{eq:46}
\end{equation}
The short-range correlations
between the two interacting protons ($p(1)$ and $p'(2)$)
and neutrons ($n(1)$ and $n'(2)$) are simulated by using
the correlation function $f(r_{12})$ in the non-antisymmetrized two-body 
matrix element (Eq. (\ref{eq:45})). We adopt its following form:
\begin{equation}
\label{eq:47}
f(r_{12})=1-e^{-\alpha r^2_{12} }(1-b r^2_{12}) \quad \mbox{with} \quad
\alpha=1.1~ \mbox{fm}^2 \quad \mbox{and} \quad  b=0.68 ~\mbox{fm}^2.
\end{equation}
The one-body transition densities entering Eq. (\ref{eq:45}) 
are given as follows:
\begin{equation}
\label{eq:48}
\frac{<J^\pi m_i\parallel [c^+_{p}{\tilde{c}}_{n}]_J\parallel 0^+_i>}
{\sqrt{2J+1}} = 
(u_{p}^{(i)} v_{n}^{(i)} {\overline{X}}^{m_i}_{(pn, J^\pi)}
+v_{p}^{(i)} u_{n}^{(i)} {\overline{Y}}^{m_i}_{(pn, J^\pi)})
\sqrt{{\cal D}^{(i)}_{pn,  J^\pi}}, 
\end{equation}
\begin{equation}
\frac{<0_f^+\parallel\widetilde{ [c^+_{p}{\tilde{c}}_{n}]_J}
\parallel J^\pi m_f> }{\sqrt{2J+1} } =
(v_{p}^{(f)} u_{n}^{(f)} 
{\overline{X}}^{m_f}_{(pn, J^\pi)}
+u_{p}^{(f)} v_{n}^{(f)} 
{\overline{Y}}^{m_f}_{(pn, J^\pi)})
\sqrt{{\cal D}^{(f)}_{pn, J^\pi}}.
\label{eq:49}
\end{equation}
The index i (f) indicates that the quasiparticles and the excited
states of the nucleus are defined with respect to the initial (final)
nuclear ground state $|0^+_i>$ ($|0^+_f>$). As 
the two sets of intermediate nuclear states generated from the 
initial and final ground states are not identical we use 
the overlap factor
\begin{eqnarray}
\label{eq:50}
<J_{m^{}_{f}}^+ | J_{m^{}_{i}}^+> &\approx & 
\sum_{p n}
\big(
\overline{X}^{m_{i}^{}}_{(pn, J^\pi)}
\overline{X}^{m_{f}^{}}_{(pn, J^\pi)}-
\overline{Y}^{m_{i}^{}}_{(pn, J^\pi)}
\overline{Y}^{m_{f}^{}}_{(pn, J^\pi)} 
\big)\times \nonumber \\ 
&& ~(u^{(i)}_{p}u^{(f)}_{p} +v^{(i)}_{p}v^{(f)}_{p}) 
(u^{(i)}_{n}u^{(f)}_{n} +v^{(i)}_{n}v^{(f)}_{n})
\end{eqnarray}
in definition (\ref{eq:45}) of the nuclear matrix elements.

%%%%%%%%%%%%%%%%%%%%%%%%%%%%%%%%%%%%%%%%%%%%%%%%%%%%%%%%%%%%%%%%%%
%%%%%%%%%%%         Calculations and discussions          %%%%%%%%
%%%%%%%%%%%%%%%%%%%%%%%%%%%%%%%%%%%%%%%%%%%%%%%%%%%%%%%%%%%%%%%%%%
\section{Results and Discussion}

%%%%%%%%%%%%%%%%%%%%%%%%%%%%%%%%%%%%%%%%%%%%%%%%%%%
\subsection{Calculation of the nuclear matrix elements}
%%%%%%%%%%%%%%%%%%%%%%%%%%%%%%%%%%%%%%%%%%%%%%%%%%%

The pn-RQRPA method has been applied for the
calculation of the SUSY $0\nu\beta\beta$-decay nuclear matrix elements
of the $A = 48$, $76$, $100$, $116$, $128$, $130$, $136$ 
and $150$ nuclear systems.
The following single particle model spaces have been considered
in these cases:\\
(i) For $^{48}Ca \rightarrow ^{48}Ti$ decay the nuclear model comprises 
13 levels:
 $0s^{}_{1/2}$,  $0p^{}_{1/2}$, $0p^{}_{3/2}$,
 $1s^{}_{1/2}$, $0d^{}_{3/2}$, $0d^{}_{5/2}$, $1p^{}_{1/2}$, $1p^{}_{3/2}$,
 $0f^{}_{5/2}$, $0f^{}_{7/2}$, $2s^{}_{1/2}$,  $0g^{}_{7/2}$, $0g^{}_{9/2}$.\\
(ii) For $^{76}Ge \rightarrow ^{76}Se$, $^{82}Se \rightarrow ^{82}Kr$
decays the model space comprises 12 levels:
 $1s^{}_{1/2}$, $0d^{}_{3/2}$, $0d^{}_{5/2}$, $1p^{}_{1/2}$, $1p^{}_{3/2}$,
 $0f^{}_{5/2}$, $0f^{}_{7/2}$, $2s^{}_{1/2}$, $1d^{}_{3/2}$, $1d^{}_{5/2}$,
 $0g^{}_{7/2}$, $0g^{}_{9/2}$. \\
(iii) For $^{96}Zr \rightarrow ^{96}Mo$, $^{100}Mo \rightarrow ^{100}Ru$ 
and $^{116}Cd \rightarrow ^{116}Sn$ decays
the model space comprises 16 levels:
 $1s^{}_{1/2}$, $0d^{}_{3/2}$, $0d^{}_{5/2}$, $1p^{}_{1/2}$, $1p^{}_{3/2}$,
 $0f^{}_{5/2}$, $0f^{}_{7/2}$, $2s^{}_{1/2}$, $1d^{}_{3/2}$, $1d^{}_{5/2}$,
 $0g^{}_{7/2}$, $0g^{}_{9/2}$, $1f^{}_{5/2}$, $1f^{}_{7/2}$, $0h^{}_{9/2}$, 
 $0h^{}_{11/2}$. \\
(iv) For $^{128}Te \rightarrow ^{128}Xe$, $^{130}Te \rightarrow ^{130}Xe$ 
and $^{136}Xe \rightarrow ^{136}Ru$ decays 
the model space comprises 16 levels:
 $1s^{}_{1/2}$, $0d^{}_{3/2}$, $0d^{}_{5/2}$, $1p^{}_{1/2}$, $1p^{}_{3/2}$,
 $0f^{}_{5/2}$, $0f^{}_{7/2}$, $2s^{}_{1/2}$, $1d^{}_{3/2}$, $1d^{}_{5/2}$,
 $0g^{}_{7/2}$, $0g^{}_{9/2}$, $2p^{}_{1/2}$, $2p^{}_{3/2}$, $1f^{}_{5/2}$, 
 $1f^{}_{7/2}$, $0h^{}_{9/2}$, $0h^{}_{11/2}$. \\
(v) For $^{150}Nd \rightarrow ^{150}Sm$ decays 
the model space comprises 20 levels:
 $1s^{}_{1/2}$, $0d^{}_{3/2}$, $0d^{}_{5/2}$, $1p^{}_{1/2}$, $1p^{}_{3/2}$,
 $0f^{}_{5/2}$, $0f^{}_{7/2}$, $2s^{}_{1/2}$, $1d^{}_{3/2}$, $1d^{}_{5/2}$,
 $0g^{}_{7/2}$, $0g^{}_{9/2}$, $2p^{}_{1/2}$, $2p^{}_{3/2}$, $1f^{}_{5/2}$, 
 $1f^{}_{7/2}$, $0h^{}_{9/2}$, $0h^{}_{11/2}$, 
 $0i^{}_{11/2}$, $0i^{}_{13/2}$. 

The single particle energies have been calculated with a 
Coulomb-corrected Wood-Saxon potential. For the two-body interaction
we used the nuclear G-matrix calculated from the Bonn one-boson 
exchange potential. The single quasiparticle energies and 
occupation amplitudes have been found by solving the BCS equations
for protons and neutrons. Since our model spaces are finite the 
proton-proton and neutron-neutron pairing interactions
have been renormalized according to Ref. \cite{cheoun}. 
In the calculation of the pn-RQRPA equation we also renormalized the 
particle--particle and particle--hole channels of the G-matrix 
interaction by introducing two parameters $g^{}_{pp}$ and $g^{}_{ph}$,
i.e. $G(pn,p'n', J) \rightarrow g_{pp} G(pn,p'n', J)$ and
$F(pn,p'n', J) \rightarrow g_{ph} F(pn,p'n', J)$.
The value adopted from our previous 
calculations \cite{FKSS97,FKS98b,SSVP97,SSF96} is $g_{{ph}} = 0.8$  
and we discuss the stability of the nuclear matrix elements in 
respect to the $g_{pp}$ inside
the expected physical range $0.8 \le g_{pp} \le 1.2$ . 

By the method outlined above we obtained the particular nuclear matrix 
elements of the two-nucleon mode ${\cal M}_{GTN}$, ${\cal M}_{FN}$, 
${\cal M}_{GT'}$ ${\cal M}_{F'}$, ${\cal M}_{T'}$ and of the pion-exchange 
mode 
${\cal M}_{GT-k\pi}$, ${\cal M}_{T-k\pi}$ for all above mentioned
nuclei.  
 Their values are listed in 
Tab. \ref{table.1} for  $g_{pp}=1.0$. 
It is worthwhile noticing that the two-nucleon  matrix 
elements are considerably smaller in comparison with the pion-exchange ones. 
From that table one can  see that the smallest nuclear matrix elements 
are associated with A=48 and 
136 systems.  We suppose that it is connected with the 
fact of the closed shell  in $^{48}$Ca and $^{136}$Xe. 
Sharp Fermi level usually offers weaker transitions due 
to the Pauli blocking effect. The largest nuclear elements are associated with 
A=150 and 100 systems. The difference between the Gamow-Teller 
matrix elements of A= 48 and A=150 systems is about an order of magnitude.  
We also stress that the one-pion 
exchange mode is disfavored in respect to the two-pion exchange mode not only 
because of a considerably smaller value of the structure coefficient 
($\alpha^{1\pi} = -0.044 \ll \alpha^{2\pi} = 0.20$) but also due to the 
partial mutual  cancellation of ${\cal M}_{GT-1\pi}$ and ${\cal M}_{T-1\pi}$
in  Eq. (\ref{eq:32}) for all studied nuclear systems. By multiplying
the particular nuclear matrix elements with corresponding structure
coefficient one obtain the full two-nucleon ${\cal M}^{2N}_{\tilde q}$,
${\cal M}^{2N}_{\tilde f}$ and pion-exchange ${\cal M}^{\pi N}$ matrix elements
[see Eqs. (\ref{eq:30})-(\ref{eq:32})]. The ${\cal M}^{2N}_{\tilde q}$ and
${\cal M}^{2N}_{\tilde f}$ have been calculated by using both 
the NR and the BM structure coefficients. 
By glancing at Tab. \ref{table.1} we can see that the values of 
${\cal M}^{2N}_{\tilde q}$ are considerably larger comparing to
${\cal M}^{2N}_{\tilde f}$ and   less sensitive to the chosen type of
structure coefficients. Also the
pion-exchange matrix elements ${\cal M}^{\pi N}$
dominate by a factor 
5 - 7 over  ${\cal M}^{2N}_{\tilde q}$ element for all 
studied nuclear systems. 

The sensitivity of the nuclear matrix elements ${\cal M}^{2N}_{\tilde q}$,
${\cal M}^{2N}_{\tilde f}$ and ${\cal M}^{\pi N}$ to the details of the
nuclear model and to the effect of short-range correlations 
are presented in
Table  \ref{table.2}. The advantage of the pn-RQRPA  in respect to the QRPA 
method is that there is no collapse of the 
pn-RQRPA solution within the physically
acceptable region of the particle-particle interaction parameter $g_{pp}$
($0.8 \le g_{pp} \le 1.2$). 
We can see that the two-nucleon matrix
element  ${\cal M}^{2N}_{\tilde f}$ is rather small and unstable in respect
to the changes of $g_{pp}$ within the discussed interval.
Moreover, its value 
crosses zero for most   nuclear systems. It means that if the deduced 
limits on $\lambda_{111}'$  associated with this nuclear matrix 
element only, the predictive power of the result would be rather small.  
The matrix elements ${\cal M}^{2N}_{\tilde f}$ 
and ${\cal M}^{\pi N}$ are suppressed
by the repulsion of the nucleon-nucleon interaction at short distances
by the  factor of about 5 and 3, respectively. Thus
the large value of  ${\cal M}^{\pi N}$ in comparison with
${\cal M}^{2N}_{\tilde f}$ 
is predominantly due to the effect of the bosonization of 
$\pi^- \to \pi^+ + 2e^-$ vertex.  We note that variations of the values of 
${\cal M}^{2N}_{\tilde f}$ and ${\cal M}^{\pi N}$ do not exceed 
$20~\%$ - $30~\%$ in respect to the  their average 
values within the allowed $g_{pp}$ interval.  Fig. \ref{fig.1}  presents 
the dependence of the two-nucleon  mode (${\cal M}^{2N}_{\tilde q}$ and 
${\cal M}^{2N}_{\tilde f}$)
and the pion-exchange mode (${\cal M}^{\pi N}$)  calculated with and without 
short range correlations
on the parameter $g_{pp}$ for A=76 system. One can see the large suppression 
of the matrix elements due to the short range correlations effect  as well as rather 
stable behavior of
the results on $g_{pp}$. The inclusion of ground state 
correlations beyond the QRPA within the pn-RQRPA method 
stabilizes the behavior of the studied matrix elements as a function 
of $g_{pp}$ and increases their predictive power, 
which is consistent with other studies \cite{SSVP97}.

%%%%%%%%%%%%%%%%%%%%%%%%%%%%%%%%%%%%%%%%%%%%%%%
\subsection{Constraints on R-parity violation}
%%%%%%%%%%%%%%%%%%%%%%%%%%%%%%%%%%%%%%%%%%%%%%%

Here we shall analyze the constraints on 
R-parity violation using the current experimental lower half-life  
limits of the $0\nu\beta\beta$-decay for the  nuclei listed in 
Table \ref{table.3}. A subject of our interest is the dependence of 
the limits of three commonly used in literature quantities
$\lambda_{111}'$, 
$\lambda_{111}'/\left( (\msq/100 \gev)^2(\mgluino/100 \gev)^{1/2}
\right)$ and 
$\lambda_{111}'/\left( (\msel/100 \gev)^2(m_{\tilde \chi^0}/100 \gev)^{1/2}
\right) $
on the GUT  scale scenarios of the MSSM. Moreover, we compare 
contributions to final limits coming from two-nucleon 
and pion--exchange modes and also 
study the relevance of the gluino and the neutralino exchange mechanisms. 

In the first step, we pay our attention to the  constraints on 
free parameters
of  MSSM,  particularly  such  SUSY  parameters as 
$\tan\beta =$ $\frac{v_2}{v_1}$, $m_0$, $m_1/2$, $A_0$ and $sign(\mu)$. 
The allowed space for these parameters is determined by 
some natural restrictions.
We use the condition that the mass eigenstates of  SUSY 
particles cannot be imaginary and consider the 
dynamical electroweak symmetry breaking 
conditions (Eq. (\ref{eq:19})). Further, we require that the  $R_7$ 
parameter 
associated  with flavor changing neutral current processes fulfills
relation (\ref{eq:22}). 
Using all the above assumptions, we have determined the excluded region for
the parameters $\tan \beta$ and $m_0$. It is shown in 
Figs. \ref{fig.2}(a)  and \ref{fig.2}(b)
where corresponding symbols indicate three main sources of constraints. 
The caption $EWSB$ indicates the points excluded by improper electroweak 
symmetry breaking, $b \to s + \gamma$ the points eliminated
by FCNC constraints and $v.e.v.$ -- the points, for which
some of the mass eigenstates become imaginary. 
One can observe that this region depends strongly on 
sgn$(\mu)$: for $\mu$ being positive,
the allowed region of $\tan \beta$ and $m_0$ is considerably 
larger. We find e.g that for $\mu > 0$ and small  values of $\tan\beta$ 
($\le 3$)  there is no restriction on $m_0$ while for $\mu < 0$  there
exists a lower bound on $m_0$ of about 240 GeV.

Further we have found that the constraints coming from the 
$B \to  X_s \gamma$ 
decay are also especially sensitive to sgn$(\mu)$. 
In Figs. \ref{fig.3}(a)-(b) 
the dependence of  the $R_7$ parameter on $m_0$ and $m_{1/2}$  is shown, 
 for both positive and negative  values of  $\mu$. 
We note that the SM contribution to $R_7$ is fixed for all the MSSM 
points at -0.188. 
The main SUSY contributions to this parameter come from charged Higgses
and charginos and are shown on Figs. \ref{fig.3}(c)-(d)
for negative  $\mu$. One can see instantly
the large negatives contribution coming  from charged Higgses 
which
is responsible for exclusion of the corresponding low values of $m_0$ 
and $m_{1/2}$.

We proceed with the determination of  limits on 
$\lambda_{111}'$, 
$\lambda_{111}'/\left( (\msq/100\gev)^2(\mgluino/100\gev)^{1/2}
\right)$ and 
$\lambda_{111}'/\left( (\msel/100\gev)^2(m_{\tilde \chi^0}/100\gev)^{1/2}
\right) $
from the non observation of double beta decay as a function of SUSY free 
parameters. For this purpose, we use  the 
$0\nu\beta\beta$-decay half-life time  (\ref{eq:29})
and the calculated nuclear matrix elements. We have found 
a very weak dependence of the quantities under discussion
 on the  $\tan\beta$ and $A_0$ SUSY
parameters ($\tan\beta$ influences mainly the Higgs (and neutralino)
sector, while $A_0$ determines a very weak sfermion mixing). 
Therefore we  present dependence  
 on $m_{1/2}$ and $m_0$ for positive $\mu$ only. 
The results are drawn in  Figs. \ref{fig.4}(a)--(c) for all nuclei we are 
interested in. 
One can see that the currently strongest
bound on these quantities is deduced from the  A=76 system followed by
A=128. The limit on the trilinear \rp breaking SUSY parameter 
$\lambda_{111}'$  
becomes less stringent with the increasing value of $m_0$ 
(Fig. \ref{fig.4}(a)). Similar 
behavior is found for 
$\lambda_{111}'/\left( (\msq/100\gev)^2(\mgluino/100\gev)^{1/2}
\right)$  also
(Fig. \ref{fig.4}(b)). On the contrary, the value of 
$\lambda_{111}'/\left( (\msel/100\gev)^2(m_{\tilde \chi^0}/100\gev)^{1/2}
\right) $
decreases with the increase of $m_0$. The reason for such a behaviour
is connected with the fact  that the mass of selectron 
is growing faster than the mass of squark 
as the value of $m_0$ is increased. In addition, the increase of 
its second power compensates for
the modest increase of $\lambda_{111}'$ in respect to $m_0$. 
This can be seen explicitely from the dependence of the  of squark,
 selectron, gluino and the lightest neutralino masses
on $m_0$ as drawn in Fig. \ref{fig.5}. We note that gluino and
neutralino are gauginos and therefore are insensitive to the  common
scalar mass at the GUT scale $m_0$. 

In Tab. \ref{table.3} we present the limits on \rp coupling
constants $\lambda_{111}'$ from the current lower bounds on the 
half-life time 
of the $0\nu\beta\beta$-decay isotopes 
experimentally most promising. 
We consider two different scales for 
$m_0$ and $m_{1/2}$: 100 GeV and 1 TeV. By glancing 
at   Tab. \ref{table.3} one finds out that the strongest limit
on $\lambda_{111}'$ comes from the $^{76}$Ge isotope and is 
$5.25\times 10^{-4}$ and $1.8\times 10^{-1}$ for the considered
values of $m_0$ and $m_{1/2}$, respectively.

The experimental constraints on the half-life of the $0\nu\beta\beta$-decay 
are expected to be more stringent in future. It would require the
recalculation of the corresponding limits on  $\lambda_{111}'$. 
For the experimentalists convenience and in order
to make our results more general we introduce the GUT constrained 
SUSY sensitivity parameter $\xi^{MSSM}_{Y}(m_0,m_{1/2})$ of a given  
isotope Y, presented in Tab. 
\ref{table.3}. This parameter incorporates the elements of both \rp SUSY and 
nuclear structure theory and is related with the limit
on $\lambda_{111}'$ as follows:
\bea
\lambda'_{111}(m_0, m_{1/2}) \le \xi^{MSSM}_Y(m_0, m_{1/2}) \times
\left(\frac{10^{24} y}{T_{1/2}^{0\nu\beta\beta -exp}}\right)^{-1/4}.
\label{eq:51}
\eea
In this way the reader is provided with an easy algorithm of 
predicting desired limits with changing the experimental data.

A detailed study of the GUT constrained SUSY scenario 
for the $0\nu\beta\beta$-decay of $^{76}Ge$ offers the most
stringent limit on R-parity breaking. At present this process is also
the most perspective  for the 
experimental detection in respect to the planned future experiment
GENIUS
\cite{Hel97}. In Figs. \ref{fig.5}(a) and \ref{fig.5}(b) we show
the  most stringent restriction on $\lambda_{111}'$ and 
$\lambda_{111}'/\left( (\msel/100\gev)^2(m_{\tilde \chi^0}/100\gev)^{1/2}
\right) $
as a function of  both $m_0$ and $m_{1/2}$. In order to be
more concrete we show  the present bounds on  $\lambda_{111}'$ together with 
those expected from the GENIUS $0\nu\beta\beta$-decay experiment
for a given set of $m_0$ and $m_{1/2}$ parameters in Tab. \ref{table.4}. 
One can  see that the 
GENIUS experiment is expected to improve the limits on  $\lambda_{111}'$ 
by about factor of 5 in respect to the current ones.  
Tab. \ref{table.4} contains also the calculated values of the 
corresponding masses of squark, selectron, gluino and lightest neutralino 
as well as the sensitivity parameter $\xi^{MSSM}_{^{76}Ge}$ within the
GUT constrained SUSY scenario. 

We also discuss the importance of different $0\nu\beta\beta$-decay 
mechanisms on hadron and quark levels.
The theoretical expression for half-life of \rp $0\nu\beta\beta$-decay in
Eq. (\ref{eq:29}) comprises the contributions from the two-nucleon mode
and pion-exchange mode, which incorporate a different combination of
\rp SUSY parameters $\eta_T$ and $\eta_{PS}$. In the previous subsection
we have shown that the pion-exchange nuclear matrix elements dominate
over the two-nucleon ones. However, it is not the necessary condition for
the dominance of pion-exchange mode. In order to clarify this point 
we calculate the limit on  $\lambda_{111}'$  by considering only
one of these mechanisms at one time.  The dependence of $\lambda_{111}'$  on 
$m_0$ is drawn  in Fig. \ref{fig.7} for $m_{1/2} = 100$ GeV
and $500$ GeV.  We see that the pion-exchange mode offers a considerably
stronger limit on  $\lambda_{111}'$ than the two-nucleon mode, 
which can be safely neglected. We note that the peak appearing
in the curves presenting the two-nucleon
mode is a consequence of interference between neutralino 
and gluino contributions to final amplitude. The curve representing
the limit on $\lambda_{111}'$ from the pion-exchange mode is free of
such instabilities. 

It remains to find out which of  the neutralino and gluino 
\rp exchange mechanisms is the most important one for the 
$0\nu\beta\beta$-decay process. In Fig. \ref{fig.8} we show 
the parameter $\lambda_{111}'$ 
as a function of $m_0$ for $m_{1/2}$ equals to 100 GeV and 500 GeV
by considering only one of the above \rp mechanisms. For
$m_{1/2}$ = 100 GeV and $m_0$ larger than about 200 GeV the gluino
mechanism is the dominant one. On the other hand,  for $m_{1/2}$ = 
500 GeV the neutralino exchange mechanism becomes prevalent 
within the whole
considered interval of $m_0$. One concludes that none of these
mechanisms is the most important  in general. The  answer
to this problem depends
on the details of the GUT constrained SUSY scenario and
in particular on the values of $m_0$ and $m_{1/2}$ parameters.

%%%%%%%%%%%%%%%%%%%%%%%%%%%%%%%%%%%%%%%%%%%%%%%%%%%%%%%%%%%%%%%%%%
%%%%%%%%%%%               Conclusions                     %%%%%%%%
%%%%%%%%%%%%%%%%%%%%%%%%%%%%%%%%%%%%%%%%%%%%%%%%%%%%%%%%%%%%%%%%%%
\section{Conclusions}

In conclusions, we have presented a detailed analysis of the 
R-parity violating trilinear terms contribution 
to the $0\nu\beta\beta$-decay within the GUT constrained MSSM
scenario. Both  nuclear and particle physics aspects of this
process have been discussed to some extent.  

We calculated relevant nuclear matrix elements within the realistic
pn-RQRPA method. A comparison of nuclear matrix elements
belonging to the two-nucleon mode and the pion-exchange mode has been performed
for A = 48, 76, 82, 96, 100, 116, 128, 130, 136 and 150 
nuclear systems. 
We have found that the $\pi$ mechanisms are larger by  
factor 5-7  mostly due to the strong
\rp $~~\pi^- \to \pi^+ + 2e^-$ transition. We suppose it is  the
explanation why the pion-exchange mode $0\nu\beta\beta$-decay 
is not favored for the mechanisms based 
on exchange of Majorana neutrinos having a different Lorentz structure
and predicting weaker \rp $~~\pi^- \to \pi^+ + 2e^-$ transition. 
Our studies show that \rp pion-exchange mode
nuclear matrix elements are less suppressed
by the short-range correlation effects and are  more stable in
respect to the details of nuclear Hamiltonian than  the two-nucleon
matrix elements. 

In the MSSM part of calculations we limited the space of the 
free SUSY parameters using necessary conditions for proper 
electroweak symmetry
breaking and inducing other limits based on current FCNC 
(Flavor Changing Neutral
Currents) experiments. The exclusion plots for the SUSY parameters
are presented and the sensitivity of the excluded region to the sign of
$\mu$ parameter is manifested. A detailed study of  the $R_7$ parameter
crucial for analysis of the FCNC $B \to X_s \gamma$ decay processes 
is also displayed and
analyzed. We also present masses of squark, selectron, gluino 
and lightest neutralino in the GUT constrained MSSM, 
the knowledge of which is required for the analysis of the \rp ~parameters. 

Using the experimental lower bounds on the $0\nu\beta\beta$-decay half-life
we then deduced current constraints on the \rp MSSM parameters for
different nuclei. The most restrictive 
constraints on the \rp Yukawa coupling constant $\lambda_{111}'$ 
are found from the $^{76}$Ge $0\nu\beta\beta$-decay experiment 
performed by the 
Heidelberg-Moscow Collaboration \cite{bau97}.
For the common SUSY masses $m_0$ and $m_{1/2}$ at the scale of 100 GeV 
and 1 TeV $\lambda_{111}'$ they are:
\begin{eqnarray}
\lambda_{111}' ~&\le &~ 5.3\times 10^{-4} ~{\rm present}, ~~~
1.1 \times 10^{-4}~({\rm GENIUS}) ~~
 {\rm for} ~~ m_0 = m_{1/2} = 100 ~\gev, \\
\lambda_{111}' ~&\le &~ 1.8\times 10^{-1} ~{\rm present}, ~~~
3.7 \times 10^{-2}~({\rm GENIUS}) ~~
 {\rm for} ~~ m_0 = m_{1/2} = 1 ~\tev. 
\end{eqnarray}
The present limit on $\lambda_{111}'$ can be improved by a factor
of 5  if the future  $0\nu\beta\beta$-decay  experiment GENIUS is
 carried out and no signal about the $0\nu\beta\beta$-decay is
detected. 

Our studies have shown that the above limits are associated with the
pion-exchange mode of the $0\nu\beta\beta$-decay, which is the dominant 
mechanism for this process at the  hadron level. Therefore,
bearing in mind the above nuclear structure analysis we argue 
that the obtained limits do
 depend very insignificantly on the nuclear physics uncertainties.

We have also dealt   with the question which of the gluino and neutralino
 $0\nu\beta\beta$-decay mechanisms is more important.
We have shown that there is no unique answer to this problem
and the dominance of any of them is bound with a
different choice of the SUSY parameters $m_0$ and $m_{1/2}$.

Finally, we conclude that the $0\nu\beta\beta$-decay imposes very 
restrictive limits on the 
important R-parity violating SUSY parameters also within the GUT
constrained MSSM scenario. The obtained constraints on the trilinear 
\rp SUSY parameter $\lambda_{111}'$ are able to compete  with those
derived from near future accelerator experiments. 

\acknowledgments

The authors acknowledge valuable discussions with S. G. Kovalenko 
on the R-parity violating MSSM. F\v S thanks the 
Department of Theoretical Physics at
Maria Curie--Sklodowska University for its hospitality. This work 
was supported in part by the State Committee for Scientific Researches
(Poland) Grant No. 2P03B00516 and
 by Grant Agency (Czech Republic) Contract No. 202/98/1216.

%%%%%%%%%%%%%%%%%%%%%%%%%%%%%%%%%%%%%%%%%%%%%%%%%%%%%%%%%%%%%%%%%%
%%%%%%%%%%%                 References                    %%%%%%%%
%%%%%%%%%%%%%%%%%%%%%%%%%%%%%%%%%%%%%%%%%%%%%%%%%%%%%%%%%%%%%%%%%%

%%%%%%%%%%%%%%%%%%%%%%%%%%%%%%%%%%%%%%%%%%%%%%%%%%%%%%%%%%%%%%%%%%
%%%%%%%%%%%              Tables                           %%%%%%%%
%%%%%%%%%%%%%%%%%%%%%%%%%%%%%%%%%%%%%%%%%%%%%%%%%%%%%%%%%%%%%%%%%%

\newpage

\widetext
\begin{table}[t]
\caption{Nuclear matrix elements (\ref{eq:30} -- \ref{eq:33}) for the 
R-parity violating SUSY mode of the $0\nu\beta\beta$-decay for the
experimentally most interesting isotopes calculated within
the renormalized pn-QRPA with $g_{pp}=1.0$. $\cal{M}\times$$10^{n}$ implies 
that the matrix element should be divided by $10^n$ to get the 
current numerical value. BM (NR) denotes that the nucleon structure
coefficients of the bag model (non--relativistic quark model)
have been considered. }
\label{table.1}
\begin{tabular}{lrrrrrrrrrr}
 & & & & & & & & & & \\
\multicolumn{11}{c}{ $(\beta\beta)_{0\nu}-decay: 
0^{+}_{g.s.}\rightarrow{0^{+}_{g.s.}}$ 
 transition} \\
 & & & & & & & & & & \\ \cline{2-11}
 & & & & & & & & & &  \\ 
M. E. & $^{48}Ca$ & $^{76}Ge$ 
& $^{82}Se$ & $^{96}Zr$ & $^{100}Mo$ &
 $^{116}Cd$ & $^{128}Te$ & $^{130}Te$ & $^{136}Xe$ & $^{150}Nd$ 
\\ 
 & & & & & & & & & & \\ \hline
\multicolumn{11}{c}{ two-nucleon mode} \\
${\cal M}_{GT-N}\times 10^2$ & 1.45 &
 7.05 & 6.52  & 4.54 & 8.14 & 4.91 & 7.37 & 6.64 & 3.87 & 11.0 \\ 
${\cal M}_{F-N}\times 10^2$ & -0.58 & 
-2.48 & -2.28 & -1.67 & -2.94 & -1.78 & -2.68 & -2.43 & -1.44 & -4.09 \\ 
${\cal M}_{GT'}\times 10^2$ & -0.21 &
-1.04 & -0.99 & -0.64 & -1.17 & -0.74 & -1.05 & -0.94 & -0.54 & -1.54 \\ 
${\cal M}_{F'}\times 10^3$ & 0.92 &
3.76 & 3.53 & 2.46 & 4.41 & 2.79 & 4.00 & 3.61 & 2.14 & 6.05 \\ 
${\cal M}_{T'}\times 10^3$ & -1.30 &
-2.38 & -1.98 & -2.68 & -4.14 & -1.88 & -3.67 & -3.46 & -1.92 & -6.34 \\ 
 & & & & & & & & &  & \\ 
${\cal M}^{2N}_{\tilde{q}}$ ~(BM) & -19.6 & 
-95.9 & -87.7 & -63.7 & -113. & -65.4 & -102. & -92.0 & -53.1 & -154. \\ 
${\cal M}^{2N}_{\tilde{f}}$ ~(BM) & -0.32 &
4.71 & 5.01 & 1.02 & 3.15 & 3.10 & 2.78 & 2.19 & 1.29 & 2.55 \\ 
 & & & & & & & & &  & \\ 
${\cal M}^{2N}_{\tilde{q}}$ ~(NR) & -27.4 &
-129. & -118.  & -85.1  & -151. & -89.3 & -137. & -123. & -71.9 & -207. \\ 
${\cal M}^{2N}_{\tilde{f}}$ ~(NR) & -1.06 &
2.22 & 2.80 & -0.89 & -0.07 & 1.28 & -0.17 & -0.52 & -0.32 & -2.13 \\ 
 & & & & & & & & &  & \\ 
\multicolumn{11}{c}{ pion mode} \\
${\cal M}_{GT-1\pi}$ & 0.25 &
1.30 & 1.23 & 0.77 & 1.43 & 0.92 & 1.25 & 1.10 & 0.61 & 1.85 \\ 
${\cal M}_{T-1\pi}$ & -0.53 &
-1.02 & -0.87 & -1.11 & -1.73 & -0.78 & -1.57 & -1.48 & -0.84 & -2.70 \\ 
${\cal M}_{GT-2\pi}$ &  -0.26 &
-1.34 & -1.26  & -0.85 & -1.52 & -0.94  & -1.40 & -1.26 & -0.74 & -2.07 \\ 
${\cal M}_{T-2\pi}$ & -0.31 &
-0.65 & -0.57 & -0.67 & -1.05 & -0.47 & -0.99 & -0.93 & -0.54 & -1.68\\ 
 & & & & & & & & &  & \\ 
${\cal M}^{\pi N}$ & -147. &
-625. & -583. & -428.  & -750.  & -436. & -692. & -627.  & -367. & -1054. \\ 
 & & & & & & & & &  & \\ 
\end{tabular}
\end{table}

\begin{table}[t]
\caption{
Sensitivity of nuclear matrix elements ${\cal M}_{\tilde q}^{2N}$, 
${\cal M}_{\tilde f}^{2N}$ and ${\cal M}^{\pi N}$
for the A=48, 76, 82, 96, 100,116,128,130,136 and 150 nuclear systems calculated 
within the renormalized pn-QRPA  to of nucleon short--range
correlations (s.r.c) and to the factor $g_{pp}$, 
renormalizing the particle-particle interaction strength. 
In  the calculations of nuclear matrix  
elements ${\cal M}_{\tilde q}^{2N}$  and ${\cal M}_{\tilde f}^{2N}$ 
within two--nucleon mode the nucleon structure coefficients of the 
non--relativistic quark model
was adopted.
}
\label{table.2}
\begin{tabular}{lrrrrrrr}
nucleus & $g_{pp}$ & \multicolumn{2}{c}{ ${\cal M}_{\tilde q}^{2N}$} &
 \multicolumn{2}{c}{ ${\cal M}_{\tilde f}^{2N}$} &
 \multicolumn{2}{c}{ ${\cal M}^{\pi N}$} \\
\cline{3-4} \cline{5-6} \cline{7-8} 
 &  & no s.r.c. & with s.r.c. & no s.r.c. & with s.r.c. & no s.r.c. & 
with s.r.c. \\ \hline
 & & & & & & & \\
$^{48}Ca$ & 0.80 & -154. & -31.2 & -22.1 & -0.85 & -440. & -164.  \\
          & 1.00 & -138. & -27.4 & -20.3 & -1.06 & -392. & -147.  \\
          & 1.20 & -119. & -23.0 & -18.1 & -1.27 & -335. & -126.  \\
 & & & & & & & \\
$^{76}Ge$ & 0.80 & -711. & -143. & -90.8 &  3.67 & -2020. & -686. \\
          & 1.00 & -646. & -128. & -80.5 &  2.22 & -1831. & -625. \\
          & 1.20 & -581. & -113. & -78.2 &  0.81 & -1645. & -564. \\
 & & & & & & & \\
$^{82}Se$ & 0.80 & -638. & -131. & -79.6 &  4.13 & -1832. & -638. \\
          & 1.00 & -581. & -119. & -74.2 &  2.80 & -1667. & -583. \\
          & 1.20 & -525. & -105. & -68.8 &  1.52 & -1503. & -529. \\
 & & & & & & & \\
$^{82}Zr$ & 0.80 & -490. & -97.0 & -66.3 &  0.25 & -1383. & -479.\\
          & 1.00 & -437. & -85.1 & -61.1 & -0.89 & -1228. & -428.\\
          & 1.20 & -358. & -73.4 & -55.9 & -1.85 & -1077. & -376. \\
 & & & & & & & \\
$^{100}Mo$ & 0.80 & -850. & -170. & -112.&  1.86 & -2408.&  -832.\\
          & 1.00  & -764. & -151. & -104.& -0.07 & -2155.& -750.\\
          & 1.20  & -678. & -131. & -95.5& -1.80 & -1908.& -668. \\
 & & & & & & & \\
$^{116}Cd$ & 0.80 & -480.& -98.2 & -61.5 & 2.05 & -1365. & -474.\\
          & 1.00 & -440. & -89.3 & -57.5 & 1.28 & -1250. & -436. \\
          & 1.20 & -401. & -80.3 & -53.5 & 0.55 & -1135. & -398.\\
 & & & & & & & \\
$^{128}Te$ & 0.80 & -778. & -155. & -103. & 1.81 & -2212.& -767.\\
          & 1.00 & -696. & -137. & -95.1 & -0.17 & -1976.& -692.\\
          & 1.20 & -615. & -118. & -87.5 & -2.08 & -1743.& -615.\\
 & & & & & & & \\
$^{130}Te$ & 0.80 & -706.& -141.& -93.6 & 1.37 &  -2008. & -697.\\
          & 1.00 & -631. & -123.& -86.7 & -0.52 & -1790. & -627.\\
          & 1.20 & -556.& -106. & -79.9 & -2.36 & -1573. & -556.\\
 & & & & & & & \\
$^{136}Xe$ & 0.80 & -418.& -83.5 & -55.0 & 1.04  & -1191.& -413.\\
          & 1.00 & -369. & -71.9 & -50.8 & -0.32 & -1048.& -367.\\
          & 1.20 & -318. & -59.7 & -46.6 & -1.74 & -901. & -317.\\
 & & & & & & & \\
$^{150}Nd$ & 0.80 & -1185. & -234. & -160. & 0.52 & -3360. &-1167. \\
          & 1.00 & -1066. & -207.& -149. & -2.13 & -3013. & -1054.\\
          & 1.20 & -949. & -180. & -138. & -4.61 & -2677. & -943. \\
\end{tabular}
\end{table}

\begin{table}[t]
\caption{Upper limits on the lepton number non-conserving parameter 
$\lambda'_{111}$ deduced from the experimental lower limits of the 
$0\nu\beta\beta$-decay half-life time $T^{0\nu\beta\beta -exp}_{1/2}(Y)$ 
for the nuclei studied in this work. The
MSSM SUSY parameters $m_0$ and $m_{1/2}$ are limited to two cases: 
100 GeV and 1 TeV. According to Eq. (51)
$\xi^{MSSM}_Y(m_0,m_{1/2})$ 
denotes  the 
sensitivity of a given nucleus Y to the $\lambda'_{111}$ parameter. 
$G_{01}$ is the integrated kinematical factor for
$0^+_{g.s.} \rightarrow 0^+_{g.s.}$ transition.}
\label{table.3}
\begin{tabular}{ccccccc}
 & & & \multicolumn{2}{c}{ $m_0 = m_{1/2} = 100 ~\gev$} &
\multicolumn{2}{c}{ $m_0 = m_{1/2} = 1 ~\tev$} \\ \cline{4-5} \cline{6-7}
nucleus & $G_{01} \times 10^{15} y$ & $T^{0\nu\beta\beta -exp}_{1/2}$(Y) [y] & 
$\xi^{MSSM}_Y$ & $\lambda'_{111}$ & $\xi^{MSSM}_Y$ & $\lambda'_{111}$ \\ \hline
$^{48}Ca$ & 803. & $9.5\times 10^{21}$ \cite{you91} &
$5.90\times 10^{-4}$ & $1.9\times 10^{-3}$ & 0.202 & 0.65 \\
$^{76}Ge$ & 7.93 & $1.1\times 10^{25}$ \cite{bau97} & 
 $9.56\times 10^{-4}$ & $5.2\times 10^{-4}$ & 0.327 & 0.18 \\
$^{82}Se$ & 35.2 & $2.7\times 10^{22}$ \cite{ell92} & 
$6.84\times 10^{-4}$ & $1.7\times 10^{-3}$ & 0.234 &  0.58 \\
$^{82}Zr$ &  73.6 & $3.9\times 10^{19}$ \cite{kaw93} & 
 $6.27\times 10^{-4}$ & $7.9\times 10^{-3}$ & 0.215  &  0.27 \\
$^{100}Mo$ &  57.3 &  $5.2\times 10^{22}$ \cite{eji96} & 
 $5.26\times 10^{-4}$ & $1.1\times 10^{-3}$ & 0.180  & 0.38 \\
$^{116}Cd$ &  62.3 & $2.9\times 10^{22}$ \cite{dane95} & 
 $6.83\times 10^{-4}$ &  $1.6\times 10^{-3}$ & 0.234 &  0.57 \\
$^{128}Te$ &  2.21 & $7.7\times 10^{24}$  \cite{bern92} & 
 $1.23\times 10^{-3}$ & $7.4\times 10^{-4}$  & 0.422  &  0.25 \\
$^{130}Te$ &  55.4 & $8.2\times 10^{21}$  \cite{ale94} & 
 $5.79\times 10^{-4}$ & $1.9\times 10^{-3}$ & 0.198 & 0.66 \\
$^{136}Xe$ &  59.1 & $4.2\times 10^{23}$ \cite{bus96} & 
 $7.44\times 10^{-4}$ & $9.2\times 10^{-4}$ & 0.254 & 0.32 \\
$^{150}Nd$ &  269. & $1.2\times 10^{21}$ \cite{sil97} & 
 $2.95\times 10^{-4}$ & $1.6\times 10^{-3}$ & 0.101 & 0.54 \\
\end{tabular}
\end{table}

\begin{table}[t]
\caption{
The supersymmetric parameters with corresponding limits on $\lambda'_{111}$
derived from the best presently available experimental limit on half-life of 
$^{76}Ge$ $0\nu\beta\beta$-decay: 
$T^{exp}_{1/2}(^{76}Ge)$  $> ~1.1\times 10^{25} ~y$
[19]. Consequences of the expected  half-life limit to be reached 
in  the GENIUS experiment  
($T^{\rm GENIUS}_{1/2}(^{76}Ge)$  $>~6\times 10^{27}$ [18])
are also shown. The $m_0$, $m_{1/2}$
$\tan(\beta)$, $A_0$ and sgn$(\mu)$ are the MSSM parameters. 
$m_{\tilde q}$, $m_{\tilde e}$, $m_{\tilde g}$ and  $m_{\chi_0^1}$
are the masses of squark, selectron, gluino and lightest neutralino,
respectively.
The $\xi^{MSSM}_{^{76}Ge}$ parameter is defined by Eq. (51).
}
\label{table.4}
\begin{tabular}{rrrrrrccc}
\multicolumn{9}{c} 
{$\tan(\beta) = 3, A_0 = 500 \gev, sign(\mu) = +1$}\\
& & & & & & & & \\ 
 $m_0$ &  $m_{1/2}$ & $~~~~m_{\tilde q}$  &
 $m_{\tilde e}$ & $m_{\tilde g}$ &
 $m_{\chi_0^1}$ &  $\xi^{MSSM}_{^{76}Ge}$   &  
$\lambda'_{111}$  & $\lambda'_{111}$  \\
 $[\gev]$ & $[\gev]$ & $[\gev]$ & $[\gev]$ & $[\gev]$ & $[\gev]$ & & 
present & GENIUS   \\ \hline
 & & & & & & &\\ 
 100   &	100  &	  250.9    &    107.2  &  261.9   &   25.8    &  
 9.56   &  $5.25\times10^{-4}$ & $1.09\times 10^{-4}$\\
 100   & 	500  &	 1136.0   &   217.7   & 1290.0   &  208.0   &  
 1.04  &  $5.73\times10^{-3}$ & $1.19\times 10^{-3}$ \\
 100   &      1000  &	 2281.0  &    399.5  &  2595.0  &   420.2  &   
 4.99  &  $2.74\times10^{-2}$ & $5.67\times 10^{-3}$\\
 500   &	100  &	   548.1  &    501.5  &     298.7  &    36.8  & 
 1.15  &   $6.31\times10^{-3}$ & $1.31\times 10^{-3}$\\
 500   &	500  &	 1237.0  &    536.1  &    1290.0  &   208.3   & 
 5.76  &  $3.16\times10^{-2}$ & $6.54\times 10^{-3}$ \\
 500   &      1000  &	2333.0  &    632.2  &   2595.0  &   420.2  &  
 1.23  &   $6.76\times10^{-2}$ & $1.40\times 10^{-2}$\\
1000  &	       100  &	1025.0  &   1001.0  &    317.7  &    39.8  &  
 4.25  &  $2.33\times10^{-2}$  & $4.83\times 10^{-3}$\\
1000  &	       500  &	1511.0   &  1019.0   & 1407.0   &  208.9   & 
 1.57  &  $8.62\times10^{-2}$ & $1.78\times 10^{-2}$\\
1000  &      1000  &	2490.0  &   1072.0  &  2596.0  &   420.3  &  
 3.27  & $1.80\times10^{-1}$ & $3.72\times 10^{-2}$\\
\end{tabular}
\end{table}

%%%%%%%%%%%%%%%%%%%%%%%%%%%%%%%%%%%%%%%%%%%%%%%%%%%%%%%%%%%%%%%%%%
%%%%%%%%%%%               Figures                         %%%%%%%%
%%%%%%%%%%%%%%%%%%%%%%%%%%%%%%%%%%%%%%%%%%%%%%%%%%%%%%%%%%%%%%%%%%
\newpage

%%%%%%%%%%%%    FIGURE 1     %%%%%%%%%%%%%%%%%%%%%%%%%%%

\begin{figure}
\caption{
The nuclear matrix elements of the two-nucleon mode
(${\cal M}^{2N}_{\tilde q}$, ${\cal M}^{2N}_{\tilde q}$) and pion-exchange
mode (${\cal M}^{\pi N}$)
of the  \rp SUSY contribution to the $0\nu\beta\beta$-decay of $^{76}$Ge
calculated within the pn-RQRPA. 
They are functions  of the particle-particle interaction strength $g_{pp}$
with (a) and without (b) 
 the two-nucleon short-range coorelations. 
}
\label{fig.1}
\end{figure}

%%%%%%%%%%%%    FIGURE 2 a, b  %%%%%%%%%%%%%%%%%%%%%%%%%%%

\begin{figure}
\caption{The constraints imposed by the $B \to  X_s \gamma$ process,
dynamical electroweak symmetry breaking condition (ESWB) and 
vacuum expectation values condition on the  $\tan \beta$ and $m_0$ SUSY
 parameters. 
The excluded region is indicated by the corresponding symbols. 
We assume $m_{1/2} = 200$ GeV  and $A_0 = 500$ GeV. 
Two cases of $\mu$ negative (a) and  $\mu$  positive (b)
are shown. 
The allowed space of parameters is considerably
larger for positive $\mu$.
}
\label{fig.2}
\end{figure}

%%%%%%%%%%%%    FIGURE 3 a,b,c,d %%%%%%%%%%%%%%%%%%%%%%%

\begin{figure}
\caption{The $R_7$ parameter  determining  the 
MSSM contributions to the flavor changing neutral current
$B \to X_s \gamma$ decays  is drawn against  
$m_0$ and $m_{1/2}$. 
The coefficient was calculated for $\tan \beta =3$, $A_0 = 500$ and
$\mu$ either positive (a) or negative (b). The charged
Higgs (c) and chargino (d) contributions to $R_7$ for
negative $\mu$ are also presented. 
}
\label{fig.3}
\end{figure}

%%%%%%%%%%%%    FIGURE 4 a,b,c    %%%%%%%%%%%%%%%%%%%%%%%%%%%

\begin{figure}
\caption
{
The limits on (a) $\lambda_{111}'$ 
 (b) $\frac{\lambda_{111}'}{(\msq/100\gev)^2(\mgluino/100\gev)^{1/2}}$ and 
(c) $\frac{\lambda_{111}'}{(\msel/100\gev)^2(m_{\tilde \chi^0}/100\gev)^{1/2}}$
deduced from the experimental lower bound on the half-life 
of the $0\nu\beta\beta$-decay of different nuclei
are plotted as a function of $m_0$. Other free parameters are fixed as follows:
$A_0 = 500$ GeV, $m_{1/2} =$ 500 GeV, 
$\tan \beta = 3$ and $\mu > 0$ (see text for details).
The needed nuclear matrix elements have been calculated within 
the pn-RQRPA.
}
\label{fig.4}
\end{figure}

%%%%%%%%%%%%    FIGURE 5  %%%%%%%%%%%%%%%%%%%%%%%%%%%

\begin{figure}
\caption{
The masses of supersymmetric particles: $m_{\tilde q}$ (squark), 
 $m_{\tilde g}$ (gluino), $m_{\tilde s}$ (selectron)  and $m_\chi$ (the lightest
neutralino)  as a function of $m_0$ within the GUT constrained minimal 
supersymmetric standard model with R-parity breaking.
Other free parameters are fixed:
 $m_{1/2} = 100$  GeV, $A_0 = 500$ GeV, $\tanb = 3$
and $\mu > 0$. 
}
\label{fig.5}
\end{figure}

%%%%%%%%%%%%    FIGURE 6 a,b  %%%%%%%%%%%%%%%%%%%%%%%%%%%

\begin{figure}
\caption{
The deduced limits on (a) $\lambda_{111}'$  and 
(b) $\frac{\lambda_{111}'}{(\msel/100\gev)^2(m_{\tilde \chi^0}/100\gev)^{1/2}}$
from the experimental lower bound on the half-life 
of the $0\nu\beta\beta$-decay in $^{76}$Ge
plotted as a function of $m_0$ and $m_{1/2}$. 
Other parameters are fixed as in the previous figures.
}
\label{fig.6}
\end{figure}

%%%%%%%%%%%%    FIGURE 7   %%%%%%%%%%%%%%%%%%%%%%%%%%%

\begin{figure}
\caption{The deduced from the  
lower bound on the half-life of the $0\nu\beta\beta$-decay 
in $^{76}$Ge limit on $\lambda_{111}'$ plotted as a function 
of $m_0$.
Two-nucleon mode (2n.-mode) and pion-exchange mode 
($\pi$-mode) are considered separately. The parameters 
$m_{1/2}$ have been chosen to be 100 GeV or 500 GeV. 
Other parameters are fixed as in the previous figures.
The dominance of the pion-exchange mode is easy to be noticed. 
}
\label{fig.7}
\end{figure}

%%%%%%%%%%%%    FIGURE 8  %%%%%%%%%%%%%%%%%%%%%%%%%%%

\begin{figure}
\caption{
The limit on $\lambda_{111}'$ deduced from the experimental lower bound 
on the half-life of the $0\nu\beta\beta$-decay 
of $^{76}Ge$ against $m_0$. 
The gluino  (g-mech.) and neutralino ($\chi$-mech.)
mechanisms in the R-parity violating scenarios
are considered separately
for two values of $m_{1/2}$ (100 GeV and
500 GeV). Other parameters are the same as in Fig. 5.
One can observe that for $m_{1/2}$ = 100 GeV and  a small value
 of $m_0$ up to 200 GeV the neutralino contribution
dominates (gives stronger constraints), while for larger values
of $m_0$ gluino mechanism becomes more important. 
}
\label{fig.8}
\end{figure}

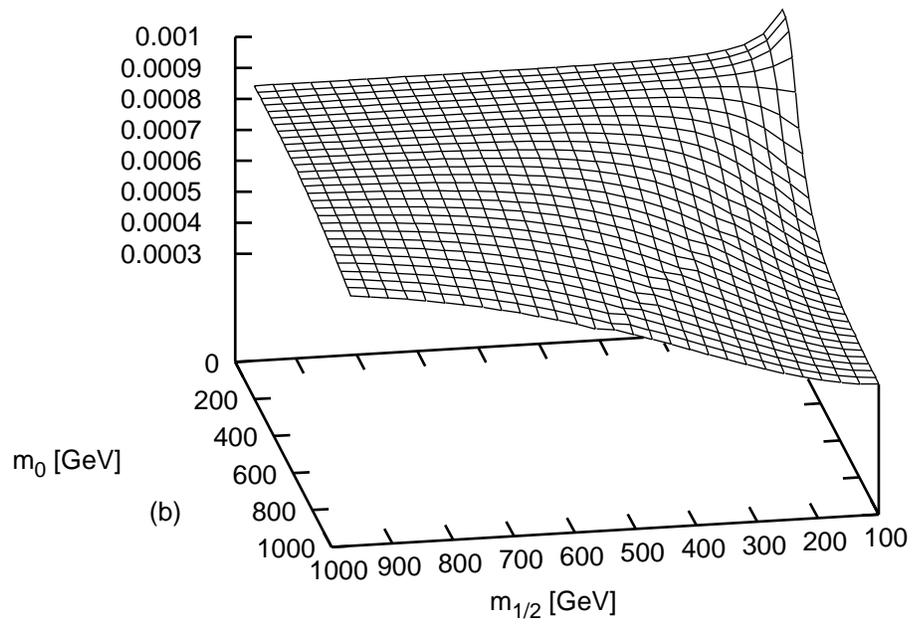
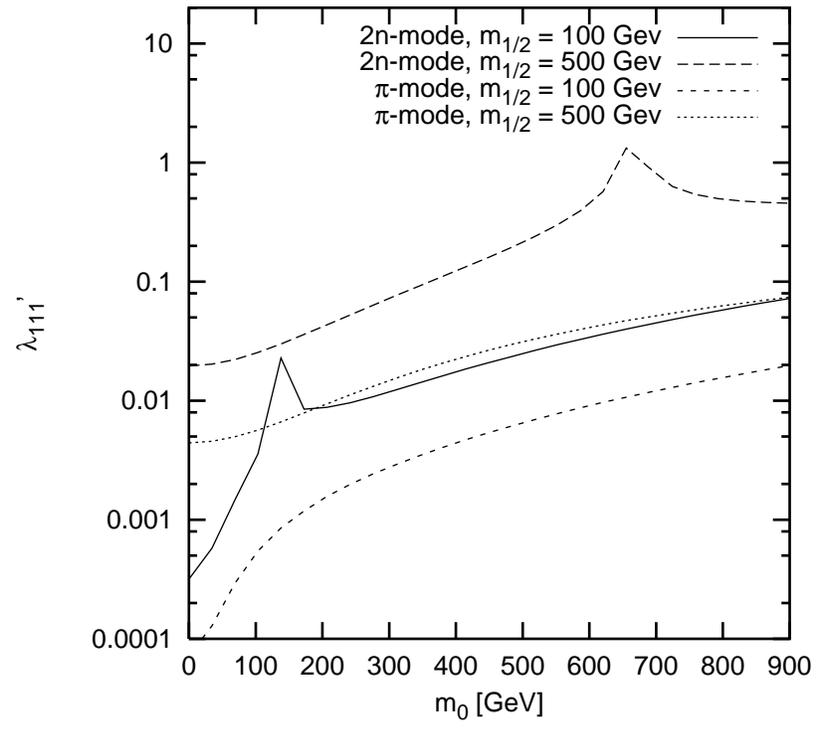
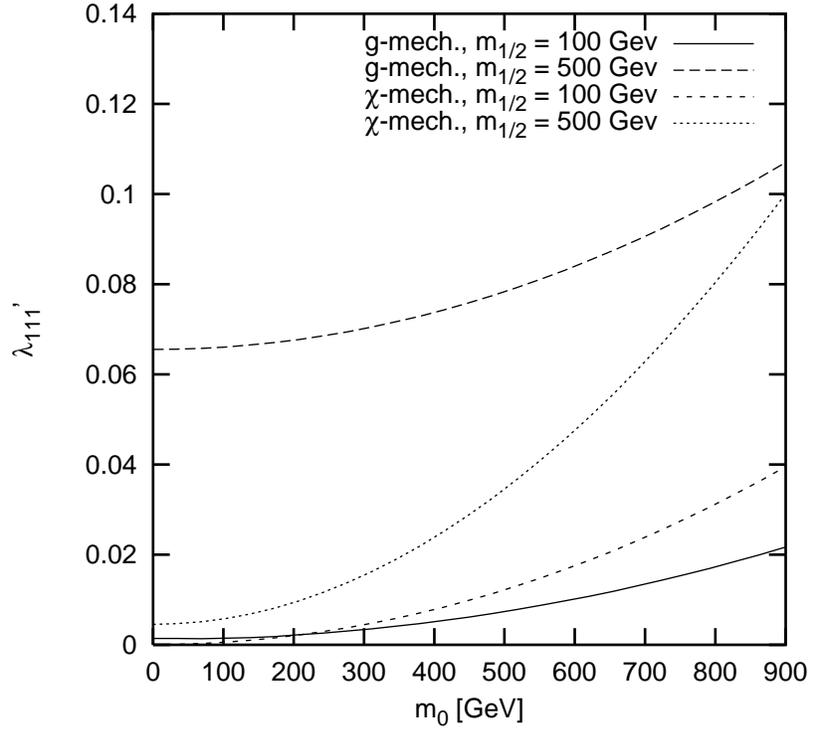
\end{document}